\documentclass[english]{IEEEtran}

\usepackage{amsthm,amsmath,amsfonts,amsfonts}
\usepackage{graphicx,color,amsmath,epsfig}
\usepackage{multicol,multirow,amssymb, cite}

\DeclareMathOperator*{\argmin}{arg\,min}

\theoremstyle{plain}
\newtheorem{theorem}{Theorem}
\newtheorem{lemma}[theorem]{Lemma}
\newtheorem{definition}[theorem]{Definition}
\newtheorem{proposition}[theorem]{Proposition}
\newtheorem{problem}[theorem]{Problem}

\newcommand{\Qed}{\hfill \ensuremath\blacksquare}

\begin{document}

\title{Blind Compressed Sensing}

\author{Sivan Gleichman and Yonina C. Eldar,~\IEEEmembership{Senior~Member,~IEEE}\thanks{This work was
supported in part by the Israel Science Foundation under Grant no.
1081/07 and by the European Commission in the framework of the FP7
Network of Excellence in Wireless COMmunications NEWCOM++
(contract no. 216715).}}
\date{}
\maketitle


\begin{abstract}
The fundamental principle underlying compressed sensing is that a signal, which is sparse under some basis representation, can be recovered from a small number of linear measurements. However, prior knowledge of the sparsity basis is essential for the recovery process. This work introduces the concept of blind compressed sensing, which avoids the need to know the sparsity basis in both the sampling and the recovery process.
We suggest three possible constraints on the sparsity basis that can be added to the problem in order to make its solution unique. For each constraint we prove conditions for uniqueness, and suggest a simple method to retrieve the solution.
Under the uniqueness conditions, and as long as the signals are sparse enough, we demonstrate through simulations that without knowing the sparsity basis our methods can achieve results similar to those of standard compressed sensing, which rely on prior knowledge of the sparsity basis.
This offers a general sampling and reconstruction system that fits all sparse signals, regardless of the sparsity basis, under the conditions and constraints presented in this work.
\end{abstract}

\section{Introduction}
Sparse signal representations have gained popularity in recent
years in many theoretical and applied areas \cite{SparseReview,
Donoho, Candes, CSforAnalog, AnalogCS, kfir}. Roughly speaking,
the information content of a sparse signal occupies only a small
portion of its ambient dimension. For example, a finite
dimensional vector is sparse if it contains a small number of
nonzero entries. It is sparse under a basis if its representation
under a given basis transform is sparse. An analog signal is
referred to as sparse if, for example, a large part of its
bandwidth is not exploited \cite{CSforAnalog, ME09T2P}. Other
models for analog sparsity are discussed in detail in
\cite{AnalogCS, kfir, EM08}.

Compressed sensing (CS) \cite{Donoho, Candes} focuses on the role
of sparsity in reducing the number of measurements needed to
represent a finite dimensional vector $x\in\mathbb{R}^m$. The
vector $x$ is measured by $b=Ax$, where $A$ is a matrix of size
$n\times m$, with $n \ll m$. In this formulation, determining $x$
from the given measurements $b$ is ill possed in general, since
$A$ has fewer rows than columns and is therefore non-invertible.
However, if $x$ is known to be sparse in a given basis $P$, then
under additional mild conditions on $A$ \cite{DonohoElad, RIP,
tropp07}, the measurements $b$ determine $x$ uniquely as long as
$n$ is large enough. This concept was also recently expanded to
include sub-Nyquist sampling of structured analog signals
\cite{CSforAnalog,kfir,MEDS09}.

In principle, recovery from compressed measurements is NP-hard.
Nonetheless, many suboptimal methods have been proposed to
approximate its solution
\cite{Donoho,Candes,SparseReview,OMP,Thresh,BP}. These algorithms
recover the true value of $x$ when $x$ is sufficiently sparse and
the columns of $A$ are incoherent \cite{SparseReview, OMP,
DonohoElad, RIP, tropp07}. However, all known recovery approaches
use the prior knowledge of the sparsity basis $P$.

Dictionary learning (DL) \cite{DicReview,miki,MIT1,MIT2,MOD} is
another application of sparse representations. In DL, we are given
a set of training signals, formally the columns of a matrix $X$.
The goal is to find a dictionary $P$, such that the columns of $X$
are sparsely represented as linear combinations of the columns of
$P$. In \cite{miki}, the authors study conditions under which the
DL problem yields a unique solution for the given training set
$X$.

In this work we introduce the concept of blind compressed sensing
(BCS), in which the goal is to recover a high-dimensional vector
$x$ from a small number of measurements, where the only prior is
that there exists some basis in which $x$ is sparse. We refer to
our setting as blind, since we do not require knowledge of the
sparsity basis for the sampling or the reconstruction. This is in
sharp contrast to CS, in which recovery necessitates this
knowledge. Our BCS framework combines elements from both CS and
DL. On the one hand, as in CS and in contrast to DL, we obtain
only low dimensional measurements of the signal. On the other
hand, we do not require prior knowledge of the sparsity basis
which is similar to the DL problem. The goal of this work is to
investigate the basic conditions under which blind recovery from
compressed measurements is possible theoretically, and to propose
concrete algorithms for this task.

Since the sparsity basis is unknown, the uncertainty about the
signal $x$ is larger in BCS than in CS. A straightforward solution
would be to increase the number of measurements. However, we show
that no rate increase can be used to determine $x$, unless the
number of measurements is equal the dimension of $x$. Furthermore,
we prove that even if we have multiple signals that share the same
(unknown) sparsity basis, as in DL, BCS remains ill-posed. In
order for the measurements to determine $x$ uniquely we need an
additional constraint on the problem. To prove the concept of BCS
we begin by discussing two simple constraints on the sparsity
basis, which enable blind recovery of a single vector $x$. We then
turn to our main contribution, which is a BCS framework for
structured sparsity bases. In this setting, we  show that multiple
vectors sharing the same sparsity pattern are needed to ensure
recovery. For all of the above formulations we demonstrate via
simulations that when the signals are sufficiently sparse the
results of our BCS methods are similar to those obtained by
standard CS algorithms which use the true, though unknown in
practice, sparsity basis. When relying on the structural
constraint we require in addition that the number of signals must
be large enough. However, the simulations show that the number of
signals needed is reasonable and much smaller than that used for
DL \cite{Unions, Unions_tech, sparseKSVD, KSVD}.

The first constraint on the basis we consider relies on the fact
that over the years there have been several bases that have been
considered "good" in the sense that they are known to sparsely
represent many natural signals. These include, for example,
various wavelet representations \cite{wavelet} and the
discrete-cosine transform (DCT) \cite{DCT}. We therefore treat the
setting in which the unknown basis $P$ is one of a finite and
known set of bases. We develop uniqueness conditions and a
recovery algorithm by treating this formulation as a series of CS
problems. To widen the set of possible bases that can be treated,
the next constraint allows $P$ to contain any sparse enough
combination of the columns of a given dictionary. We show that the
resulting CS problem can be viewed within the framework of
standard CS, or as DL with a sparse dictionary \cite{sparseKSVD}.
We compare these two approaches for BCS with a sparse basis. For
both classes of constrains we show that a Gaussian random
measurement matrix satisfies the uniqueness conditions we develop
with probability one.

Our main contribution is inspired by multichannel systems, where
the signals from each channel are sparse under separate bases. In
our setting this translates to the requirement that $P$ is block
diagonal. For simplicity, and following several previous works
 \cite{OrthMix, DicIdent, DicIdentUni}, we impose in addition that $P$
is orthogonal. We then choose to measure the set of signals $X$ by
a measurement matrix $A$ consisting of a union of orthogonal
bases. This choice has been used in previous CS and DL works as
well \cite{UncerPrinc, UncerPrincPair,
SparseRepUnion,Unions,Unions_tech}. For technical reasons we also
choose the number of blocks in $P$ as an integer multiple of the
number of bases in $A$. Using this structure we develop uniqueness
results as well as a concrete recovery algorithm. The uniqueness
condition follows from reformulating the BCS problem within the
framework of DL and then relying on results obtained in that
context. In particular, we require an ensemble of signals $X$, all
sparse in the same basis. As we show, a suitable choice of random
matrix $A$ satisfies the uniqueness conditions with probability~1.

Unfortunately, the reduction to an equivalent DL problem which is
used for the uniqueness proof, does not lead to a practical
recovery algorithm. This is due to the fact that it necessitates
resolving the signed permutation ambiguity, which is inherent in
DL. Instead, we propose a simple and direct algorithm for
recovery, which we refer to as the orthogonal block diagonal BCS
(OBD-BCS) algorithm. This method finds $X=PS$ by computing a basis
$P$ and a sparse matrix $S$ using two alternating steps. The first
step is sparse coding, in which $P$ is fixed and $S$ is updated
using a standard CS algorithm. In the second step $S$ is fixed and
$P$ is updated using several singular value decompositions (SVD).

The remainder of the paper is organized as follows. In
Section~\ref{sec:ProbDef} we review the fundamentals of CS and
define the BCS problem. In Section~\ref{sec:Unique} we prove that
BCS is ill posed by showing that it can be interpreted as a
certain ill-posed DL problem. In
Sections~\ref{sec:finite},~\ref{sec:sparse},~\ref{sec:struct} we
consider the three constrained BCS problems respectively. A
comparison between the different approaches is provided in
Section~\ref{sec:SimComp}.


\section{BCS Problem Definition}\label{sec:ProbDef}

\subsection{Compressed Sensing}\label{sec:CS}
We start by shortly reviewing the main results in the field of CS
needed for our derivations. The goal of CS is to reconstruct a
vector $x\in \mathbb{R}^m$ from measurements $b=Ax$, where
$A\in\mathbb{R}^{n\times m}$ and $n \ll m$. This problem is ill
possed in general and therefore has infinitely many possible
solutions. In CS we seek the sparsest solution:
\begin{equation}\label{P0x}
    \hat{x}=\argmin{||x||_0} \qquad \text{s.t.} \qquad b=Ax,
\end{equation}
where $||\cdot||_0$ is the $\ell_0$ semi-norm which counts the number of nonzero elements of the vector. This idea can be generalized to the case in which $x$ is sparse under a given basis $P$, so that there is a sparse vector $s$ such that $x=Ps$. Problem \eqref{P0x} then becomes
\begin{equation}\label{P01}
    \hat{s}=\argmin{||s||_0} \qquad \text{s.t.} \qquad b=APs,
\end{equation}
and the reconstructed signal is $\hat{x}=P\hat{s}$. When the
maximal number of nonzero elements in $s$ is known to equal $k$,
we may consider the objective
\begin{equation}\label{P0}
    \hat{s}=\argmin{||b-APs||_2^2} \qquad \text{s.t.} \qquad ||s||_0\leq k.
\end{equation}

An important question is under what conditions \eqref{P0x}-\eqref{P0} have a unique solution.
In \cite{DonohoElad} the authors define the \emph{spark} of a matrix, denoted by $\sigma(\cdot)$, which is the smallest possible number of linearly dependent columns. They prove that if $s$ is $k$-sparse, and $\sigma(AP)\geq2k$, then the solution to \eqref{P01}, or equivalently \eqref{P0}, is unique.
Unfortunately, calculating the spark of a matrix is a combinatorial problem. However, it is often bounded by the \emph{mutual coherence} \cite{DonohoElad}, which can be calculated easily. Denoting the $i$th column of a matrix $D$ by $d_i$, the mutual coherence of $D$ is given by
\begin{equation*}
    \mu(D)=\max_{i \neq j}{\frac{|d_i^Td_j|}{||d_i||_2||d_j||_2}}.
\end{equation*}
It is easy to see that $\sigma(D)\geq1+\frac{1}{\mu(D)}$. Therefore, a sufficient condition for the uniqueness of the solutions to \eqref{P01} or \eqref{P0} is
\begin{equation*}
    k\leq\frac{1}{2}\left(1+\frac{1}{\mu(AP)}\right).
\end{equation*}

Although the uniqueness condition involves the product $AP$, some
CS methods are universal. This means that by constructing a
suitable measurement matrix $A$, uniqueness is guaranteed for any
fixed orthogonal basis $P$. In such cases knowledge of $P$ is not
necessary for the sampling process. One way to achieve this
universality property with probability~1 relies on the next
proposition.
\begin{proposition} \label{prop:spark}
    If $A$ is an i.i.d. Gaussian random matrix of size $n\times m$, where $n<m$, then $\sigma(AP)=n+1$ with probability~1 for any fixed orthogonal basis $P$.
\end{proposition}

\emph{Proof}:
Due to the properties of Gaussian random variables and since $P$ is orthogonal, the product $AP$ is also an i.i.d. Gaussian random matrix. Since any $n$, or less, i.i.d. Gaussian vectors in $\mathbb{R}^n$ are linearly independent with probability~1, $\sigma(AP)>n$ with probability~1. On the other hand, more then $n$ vectors in $\mathbb{R}^n$ are always linearly dependent, therefore $\sigma(AP)= n+1$.
\Qed

According to Proposition~\ref{prop:spark} if $A$ is an i.i.d Gaussian matrix and the number of nonzero elements in $s$ is $k\leq n/2$, then the uniqueness of the solution to \eqref{P01} or \eqref{P0} is guaranteed with probability~1 for any fixed orthogonal basis $P$ (see also \cite{RandomRIP}).

Problems \eqref{P01} and \eqref{P0} are NP-hard in general. Many
suboptimal methods have been proposed to approximate their
solutions, such as
\cite{Donoho,Candes,SparseReview,OMP,Thresh,BP}. These algorithms
can be divided into two main approaches: greedy algorithms and
convex relaxation methods. Greedy algorithms approximate the
solution by selecting the indices of the nonzero elements in
$\hat{s}$ sequentially. One of the most common methods of this
type is orthogonal matching pursuit (OMP) \cite{OMP}. Convex
relaxation approaches change the objective in \eqref{P01} to a
convex problem. The most common of these methods is basis pursuit
(BP) \cite{BP}, which considers the problem:
\begin{equation}\label{P1}
    \hat{s}=\argmin{||s||_1} \qquad \text{s.t.} \qquad b=APs.
\end{equation}
Under suitable conditions on the product $AP$ and the sparsity
level of the signals, both the greedy algorithms and the convex
relaxation methods recover the true value of $s$. For instance,
both OMP and BP recover the true value of $s$ when the number of
nonzero elements in $s$ is no more than
$\frac{1}{2}(1+\frac{1}{\mu(AP)})$ \cite{SparseReview, OMP,
DonohoElad, RIP, tropp07}.

\subsection{BCS Problem Formulation}
Even when the universality property is achieved in CS, all existing algorithms require the knowledge of the sparsity basis $P$ for the reconstruction process. The idea of BCS is to avoid entirely the need of this prior knowledge. That is, perform both the sampling and the reconstruction of the signals without knowing under which basis they are sparse.

This problem seems impossible at first, since every signal is sparse under a basis that contains the signal itself. This would imply that BCS allows reconstruction of any signal from a small number of measurements without any prior knowledge, which is clearly impossible.
Our approach then, is to sample an ensemble of signals that are all sparse under the same basis. Later on we revisit problems with only one signal, but with additional constraints.

Let $X\in \mathbb{R}^{m \times N} $ denote a matrix whose columns are the original signals, and let $S \in \mathbb{R}^{m \times N} $ denote the matrix whose columns are the corresponding sparse vectors, such that $X=PS$ for some basis $P\in \mathbb{R}^{m \times m}$. The signals are all sampled using a measurement matrix $A\in \mathbb{R}^{n \times m}$, producing the matrix $B=AX$. For the measurements to be compressed the dimensions should satisfy $n<m$, where the compression ratio is $L=m/n$. Following \cite{miki, KSVD} we assume the maximal number of nonzero elements in each of the columns of $S$, is known to equal $k$. We refer to such a matrix $S$ as a $k$-sparse matrix.
The BCS problem can be formulated as follows.
\begin{problem}\label{prob:original}
    Given the measurements $B$ and the measurement matrix $A$ find the signal matrix $X$ such that $B=AX$ where $X=PS$ for some basis $P$ and $k$-sparse matrix $S$.
\end{problem}

Note that our goal is not to find the basis $P$ and the sparse matrix $S$. We are only interested in the product $X=PS$. In fact, for a given matrix $X$ there is more than one pair of matrices $P$ and $S$ such that $X=PS$. Here we focus on the question of whether $X$ can be recovered given the knowledge that such a pair exists for $X$.

\section{Uniqueness}\label{sec:Unique}
We now discuss BCS uniqueness, namely the uniqueness of the signal matrix $X$ which solves  Problem~\ref{prob:original}.
Unfortunately, although Problem~\ref{prob:original} seems quite natural, its solution is not unique for any choice of measurement matrix $A$, for any number of signals and any sparsity level.
We prove this result by reducing the problem to an equivalent one, using the field of DL, and proving that the solution to the equivalent problem is not unique.

In Section~\ref{sec:DL} we review results in the field of DL needed for our derivation. In Section~\ref{sec:new_prob} we use these results to prove that the BCS problem does not have a unique solution.
In Sections~\ref{sec:finite},~\ref{sec:sparse},~\ref{sec:struct} we suggest several constraints on the basis $P$ that ensure uniqueness.

\subsection{Dictionary Learning (DL)}\label{sec:DL}
The field of DL \cite{DicReview,miki,MIT1,MIT2,MOD} focuses on
finding a sparse matrix $S\in\mathbb{R}^{m\times N}$ and a
dictionary $D\in\mathbb{R}^{n\times m}$ such that $B=DS$ where
only $B\in\mathbb{R}^{n\times N}$ is given. Usually in DL the
dimensions satisfy $n \ll m$. BCS can be viewed as a DL problem
with $D=AP$ where $A$ is known and $P$ is an unknown basis. Thus,
one may view BCS as a DL problem with a constrained dictionary.
However, there is an important difference in the output of DL and
BCS. DL provides the dictionary $D=AP$ and the sparse matrix $S$.
On the other hand, in BCS we are interested in recovering the
unknown signals $X=PS$. Therefore, after performing DL some
postprocessing is needed to retrieve $P$ from $D$. This is an
important distinction which, as we show in
Section~\ref{sec:OBDalg}, makes it hard to directly apply DL
algorithms.

An important question is the uniqueness of the DL factorization. That is, given a matrix $B\in \mathbb{R}^{n \times N}$ what are the conditions for the uniqueness of the pair of matrices $D\in \mathbb{R}^{n \times m}$ and $S\in \mathbb{R}^{m \times N}$ such that $B=DS$ where $S$ is $k$-sparse. Note that if some pair $D,S$ satisfies $B=DS$, then scaling and signed permutation of the columns of $D$ and rows of $S$ respectively do not change the product $B=DS$. Therefore, there cannot be a unique pair $D,S$. In the context of DL the term uniqueness refers to uniqueness up to scaling and signed permutation.
In fact in most cases without loss of generality we can assume the columns of the dictionary have unit norm, such that there is no ambiguity in the scaling, but only in the signed permutation.

Conditions for DL uniqueness when the dictionary $D$ is orthogonal or just square are provided in \cite{DicIdent} and \cite{DicIdentUni}. However, in BCS $D=AP$ is in general rectangular. In \cite{miki} the authors prove sufficient conditions on $D$ and $S$ for the uniqueness of a general DL. We refer to the condition on $D$ as the \emph{spark condition} and to the conditions on $S$ as the \emph{richness conditions}. The main idea behind these conditions is that $D$ should satisfy the condition for CS uniqueness, and that the columns of $S$ should be diverse regarding both the locations and the values of the nonzero elements. More specifically, the conditions for DL uniqueness are:

\begin{itemize}
    \item The spark condition: $\sigma(D)\geq 2k$.
    \item The richness conditions:
    \begin{enumerate}
        \item All the columns of $S$ have exactly $k$ nonzero elements.
        \item For each possible $k$-length support there are at least $k+1$ columns in $S$.
        \item Any $k+1$ columns in $S$, which have the same support, span a $k$-dimensional space.
        \item Any $k+1$ columns in $S$, which have different supports, span a $(k+1)$-dimensional space.
        \end{enumerate}
\end{itemize}

According to the second of the richness conditions the number of signals, that is the number of columns in $S$, must be at least $\binom{m}{k}(k+1)$. Nevertheless, it was shown in \cite{miki} that in practice far fewer signals are needed. Heuristically, the number of signals should grow at least linearly with the length of the signals.
It was also shown in \cite{miki} that DL algorithms perform well even when there are at most $k$ nonzero elements in the columns of $S$ instead of exactly $k$.

\subsection{BCS Uniqueness}\label{sec:new_prob}
Under the conditions above the DL solution given the measurements
$B$ is unique. That is, up to scaling and signed permutations
there is a unique pair $D,S$ such that $B=DS$ and $S$ is
$k$-sparse. Since we are interested in the product $PS$ and not in
$P$ or $S$ themselves, without loss of generality we can always
assume that the columns of $P$ are scaled so that the columns of
$D=AP$ have unit norm. This way there is no ambiguity in the
scaling of $D$ and $S$, but only in their signed permutation. That
is, applying DL on $B$ provides $\tilde{D}=APQ$ and
$\tilde{S}=Q^TS$ for some unknown signed permutation matrix $Q$. A
signed permutation matrix is a column (or row) permutation of the
identity matrix, where the sign of each column (or row) can change
separately. In other words, it has only one nonzero element, equal
$\pm1$, in each column and each row. Any signed permutation matrix
is obviously orthogonal.

If we can find the basis $\tilde{P}=PQ$ out of $\tilde{D}$, then we can recover the correct signal matrix by:
\begin{equation*}
\tilde{P}\tilde{S}=PQQ^TS=PS=X.
\end{equation*}
Therefore, under the uniqueness conditions for DL on $S$ and $D=AP$ Problem~\ref{prob:original} is equivalent to the following problem.

\begin{problem}\label{prob:new}
Given $\tilde{D}\in\mathbb{R}^{n\times m}$ and $A\in\mathbb{R}^{n\times m}$, where $n<m$, find a basis $\tilde{P}$ such that $\tilde{D}=A\tilde{P}$.
\end{problem}

We therefore focus on the uniqueness of Problem~\ref{prob:new}.
Since $n<m$ the matrix $A$ has a null space. As we now show, even with the constraint that $\tilde{P}$ is a basis there is still no unique solution.

To see that assume $\tilde{P}_1$ is a basis, i.e., has full rank, and satisfies $\tilde{D}=A\tilde{P}_1$. Decompose $\tilde{P}_1$ as $\tilde{P}_1=P_{N^{\bot}}+P_N$ where the columns of $P_N$ are in $N(A)$, the null space of $A$, and those of $P_{N^{\bot}}$ are in its orthogonal complement $N(A)^{\bot}$. Note that necessarily $P_N\neq0$, otherwise the matrix $\tilde{P}_1=P_{N^{\bot}}$ is in $N(A)^{\bot}$ and has full rank. However, since the dimension of $N(A)^{\bot}$ is at most $n<m$, it contains at most $n$ linearly independent vectors. Therefore, there is no $m \times m$ full rank matrix whose columns are all in $N(A)^{\bot}$.

Next define the matrix $\tilde{P}_2=P_{N^{\bot}}-P_N$ which is different from $\tilde{P}_1$, but it is easy to see that $\tilde{D}=A\tilde{P}_2$. Moreover, since the columns of $P_N$ are perpendicular to the columns of $P_{N^{\bot}}$,
\begin{equation*}
\tilde{P}_1^T\tilde{P}_1=\tilde{P}_2^T\tilde{P}_2=||P_{N^{\bot}}||_F^2+||P_N||_F^2.
\end{equation*}
A square matrix $P$ has full rank if and only if $P^TP$ has full rank. Therefore, since $\tilde{P}_1$ has full rank and $\tilde{P}_2^T\tilde{P}_2=\tilde{P}_1^T\tilde{P}_1$, $\tilde{P}_2$ also has full rank. So that both $\tilde{P}_1$ and $\tilde{P}_2$ are solutions to Problem~\ref{prob:new}.
In fact there are many more solutions; some of them can be found by changing the signs of only part of the columns of $P_N$.

We now return to the original BCS problem, as defined in
Problem~\ref{prob:original}. We just proved that when the DL
solution given $B$ is unique, Problem~\ref{prob:original} is
equivalent to Problem~\ref{prob:new} which has no unique solution.
Obviously if the DL solution given $B$ is not unique, then BCS
will not be unique. Therefore, Problem~\ref{prob:original} has no
unique solution for any choice of parameters.

In order to guarantee a unique solution we need an additional constraint. We next discuss constraints on $P$ that can render the solution to Problem~\ref{prob:new} unique, and therefore in addition to the richness conditions on $S$ and the spark condition on $AP$ they guarantee the uniqueness of the solution to Problem~\ref{prob:original}.
Although there are many possible constraints, we focus below on the following.
\begin{enumerate}
    \item $P$ is one of a finite and known set of bases.
    \item $P$ is sparse under some known dictionary.
    \item $P$ is orthogonal and has a block diagonal structure.
\end{enumerate}

The motivation for these constraints comes from the uniqueness of
Problem~\ref{prob:new}. Nonetheless, we provide conditions under
which the solution to Problem~\ref{prob:original} with constraints
1 or 2 is unique even without DL uniqueness. In fact, under these
conditions the solution to Problem~\ref{prob:original} is unique
even when $N=1$, so that there is only one signal.

In the next sections we consider each one of the constraints, prove conditions for the uniqueness of the constrained BCS solution, and suggest a method to retrieve the solution. Table~\ref{tab:summary} summarizes these three approaches.

\begin{table*}[tbp]
\caption{summary of constraints on $P$}\label{tab:summary}
\begin{center}
\begin{tabular}{|l|l|l|}
\hline
\multicolumn{1}{|c|}{\textbf{The constraint}} & \multicolumn{1}{|c|}{\textbf{Conditions for uniqueness}} & \multicolumn{1}{|c|}{\textbf{Algorithm}} \\

\hline
\textbf{Finite Set} - Section~\ref{sec:finite} & $\bullet\; \sigma(AP)\leq 2k$ for any $P\in\Psi$. & $\bullet\;$ F-BCS - Solving \eqref{eq:FBCSsingel_1} or \eqref{eq:FBCSsingel_2} for each $P\in\Psi$ using a standard CS \\
$P$ is in a given finite set& $\bullet\; A$ is $k$-rank preserving of $\Psi$ (Definition~\ref{def:rank}). &  $\;\;\;$ algorithm, and choosing the best solution.\\
of possible bases $\Psi$. & & \\

\hline
\textbf{Sparse Basis} - Section~\ref{sec:sparse} & $\bullet\;\sigma(A\Phi)\geq2k_Pk$. & $\bullet\;$ Direct method - Solving \eqref{eq:Sparse_P01} or \eqref{eq:Sparse_P0} using a standard CS algorithm,\\
$P$ is $k_P$-sparse under a& &  $\;\;\;$ where the recovery is $X=\Phi C$. \\
given dictionary $\Phi$. && $\bullet\;$ Sparse K-SVD - Using sparse K-SVD algorithm \cite{sparseKSVD} to retrieve $S,Z$,\\
& & $\;\;\;$ where the recovery is $X=\Phi ZS$.\\

\hline
\textbf{Structure} - Section~\ref{sec:struct} & $\bullet\;$The richness conditions on $S$. & $\bullet\;$ OBD-BCS - Updating $S$ and $P$ alternately according to the algorithm\\
$P$ is orthogonal $2L$-block & $\bullet\;A$ is a union of $L$ orthogonal bases. &$\;\;\;$ in Table~\ref{tab:OBD}, where the recovery is $X=PS$. \\
diagonal. & $\bullet\;\sigma(AP)=n+1$. &\\
& $\bullet\;A$ is not inter-block diagonal (Definition~\ref{def:InterBlock}). &\\
\hline
\end{tabular}
\end{center}
\end{table*}

\section{Finite Set of Bases}\label{sec:finite}
One way to guarantee a unique solution to Problem~\ref{prob:new} is to limit the number of possible bases $\tilde{P}$ to a finite set of bases, and require that these bases are different from one another under the measurement matrix $A$.
Since $\tilde{P}$ in Problem~\ref{prob:new} is a column signed permutation of $P$ in Problem~\ref{prob:original}, by limiting $P$ to a finite set we also limit the possible $\tilde{P}$ to a finite set.
The new constrained BCS, instead of Problem~\ref{prob:original}, is then:
\begin{problem} \label{prob:finite}
    Given the measurements $B$, the measurement matrix $A$ and a finite set of bases $\Psi$, find the signal matrix $X$ such that $B=AX$ and $X=PS$ for some basis $P\in\Psi$ and $k$-sparse matrix $S$.
\end{problem}

The motivation behind Problem~\ref{prob:finite} is that over the
years a variety of bases were proven to lead to sparse
representations of many natural signals, such as wavelet
\cite{wavelet} and DCT \cite{DCT}. These bases have fast
implementations and are known to fit many types of signals.
Therefore, when the basis is unknown it is natural to try one of
these choices.

\subsection{Uniqueness Conditions}
We now show that under proper conditions the solution to Problem~\ref{prob:finite} is unique even when there is only one signal, namely $N=1$. In this case instead of the matrices $X,S,B$ we deal with the vectors $x,s,b$ respectively.

Assume $x$ is a solution to Problem~\ref{prob:finite}. That is,
$x$ is $k$-sparse under $P\in\Psi$ and satisfies $b=Ax$.
Uniqueness is achieved if there is no $\bar{x}\neq x$ which is
$k$-sparse under a basis $\bar{P}\in\Psi$ and also satisfies
$b=A\bar{x}$. We first require that $\sigma(AP)\geq 2k$; otherwise
even if $\bar{P}=P$ there is no unique solution \cite{DonohoElad}.
Since the real sparsity basis $P$ is unknown we require that
$\sigma(AP)\geq 2k$ for any $P\in\Psi$.

Next we write $x=Ps=P_Ts_T$, where $T$ is the index set of the
nonzero elements in $s$ with $|T|\leq k$, $s_T$ is the vector of
nonzero elements in $s$, and $P_T$ is the sub-matrix of $P$
containing only the columns with indices in $T$. If $\bar{x}$ is
also a solution to Problem~\ref{prob:finite} then
$\bar{x}=\bar{P}\bar{s}=\bar{P}_{J}\bar{s}_{J}$, where $J$ is the
index set of the nonzero elements in $\bar{s}$, and $|J|\leq k$.
Moreover, $b=A\bar{P}_{J}\bar{s}_{J}=AP_Ts_T$, which implies that
the matrix $A[P_T,\bar{P}_J]$ has a null space. This null space
contains the null space of $[P_T,\bar{P}_J]$. By requiring
\begin{equation}\label{eq:rank}
\text{rank}(A[P_T,\bar{P}_J])=\text{rank}[P_T,\bar{P}_J],
\end{equation}
we guarantee that the null space of $A[P_T,\bar{P}_J]$ equals the null space of $[P_T,\bar{P}_J]$. Therefore, under \eqref{eq:rank}, $A\bar{P}_{J}\bar{s}_{J}=AP_Ts_T$ if and only if $\bar{P}_{J}\bar{s}_{J}=P_Ts_T$, which implies $\bar{x}=x$.

Therefore, in order do guarantee the uniqueness of the solution to Problem~\ref{prob:finite} in addition to the requirement that $\sigma(AP)\geq 2k$ for any $P\in\Psi$, we require that any two index sets $T,J$ of size $k$ and any two bases $P,\bar{P}\in\Psi$ satisfy \eqref{eq:rank}.
\begin{definition}\label{def:rank}
A measurement matrix $A$ is \emph{$k$-rank preserving of the bases set $\Psi$} if any two index sets $T,J$ of size $k$ and any two bases $P,\bar{P}\in\Psi$ satisfy \eqref{eq:rank}.
\end{definition}

The conditions for the uniqueness of the solution to Problem~\ref{prob:finite} are therefore: $\sigma(AP)\geq 2k$ for any $P\in\Psi$, and $A$ is $k$-rank preserving of the set $\Psi$.
In order to satisfy the first condition with probability~1, according to Section~\ref{sec:CS} we can require all $P\in\Psi$ to be orthogonal and generate $A$ from an i.i.d. Gaussian distribution. However, since the number of bases is finite, we can instead verify the first condition is satisfied by checking the spark of all the products $AP$. Alternatively, one can bound the spark of these matrices using their mutual coherence.

It is easy to see that any full column rank matrix $A$ is $k$-rank
preserving for any $k$ and any set $\Psi$. However, in our case
$A$ is rectangular and therefore does not have full column rank.
In order to guarantee that $A$ is $k$-rank preserving with
probability~1 we rely on the following proposition:
\begin{proposition}
    An i.i.d Gaussian matrix $A$ of size $n\times m$ is with probability~1 $k$-rank preserving of any fixed finite set of bases and any $k\leq n/2$.
\end{proposition}

\emph{Proof}: If $n\geq m$ then $A$ has full column rank with probability~1, and is therefore $k$-rank preserving with probability~1. We therefore focus on the case where $n<m$. Assume $T,J$ are index sets of size $k$, and $P,\bar{P}\in\Psi$.
Denote $r=\text{rank}[P_T,\bar{P}_J]$. We then need to prove that $\text{rank}(A[P_T,\bar{P}_J])=r$.

Perform a Gram Schmidt process on the columns of $[P_T,\bar{P}_J]$ and denote the resulting matrix by $G$. $G$ is then an $m\times r$ matrix with orthonormal columns, with $\text{rank}(G)=r$ and $\text{rank}(AG)=\text{rank}(A[P_T,\bar{P}_J])$.
Next we complete $G$ to an orthogonal matrix $G_u$ by adding columns. According to Proposition~\ref{prop:spark} since $A$ is an i.i.d Gaussian matrix and $G_u$ is orthogonal $\sigma(AG_u)=n+1$ with probability~1. Therefore, with probability~1 any $t$ columns of $AG_u$ are linearly independent, with $t\leq n$. In particular, with probability~1 the columns of $AG$ are linearly independent, so that $\text{rank}(AG)=r$, completing the proof.
\Qed

Until now we proved conditions for the uniqueness of Problem~\ref{prob:finite} when there is only one signal $N=1$. The same conditions are true for $N>1$ since we can look at every signal separately. However, since all the signals are sparse under the same basis, if $N>1$ then the condition that $A$ must be $k$-rank preserving can be relaxed.

For instance, consider the case where there are only two index sets $T,J$ and two bases $P,\bar{P}\in\Psi$ ($P$ is the real sparsity basis) that do not satisfy \eqref{eq:rank}.
In this case if we have many signals with different sparsity patterns, then only a small portion of them fall in the problematic index set, and therefore might falsely indicate that $\bar{P}$ is the sparsity basis. However, most of the signals correspond to index sets that satisfy \eqref{eq:rank}, and therefore these signals indicate the correct basis. The selection of the sparsity bases is done according to the majority of signals and therefore the correct basis is selected.

Another example is the case where there are enough diverse signals such that the richness conditions on $S$ are satisfied. In this case it is enough to require that for any two bases $P,\bar{P}\in\Psi$ the matrices $AP$ and $AP$ are different from one another even under scaling and signed permutation of the columns. This way we guarantee that the problem equivalent to Problem~\ref{prob:finite} under the richness and spark conditions has a unique solution, and therefore Problem~\ref{prob:finite} also has a unique solution.

Problem~\ref{prob:finite} can also be viewed as a CS problem with a block sparsity constraint \cite{BlockSparsity1, BlockSparsity2}. That is, if $\Psi=\{P_1,P_2,...\}$ then the desired signal matrix can be written as
\begin{equation*}
    X=[P_1,P_2,...]\left[\begin{array}{ccc}S_1\\ S_2\\ \vdots \end{array}\right],
\end{equation*}
where only one of the submatrices $S_i$ is not all zeros.
In contrast to the usual block sparsity constraint here the sub-matrix $S_i$ which is not zero is itself sparse.
However, the uniqueness conditions which are implied from this block sparsity CS approach are too strong comparing to our BCS approach. For instance, they require all $P_j\in\Psi$, to be incoherent, whereas the BCS uniqueness is not disturbed by coherent bases. In fact the solution is unique even if the bases in $\Psi$ equal one another.
This is because here we are not interested in recovering $S_i$ but rather $P_iS_i$.

\subsection{The F-BCS Method}
The uniqueness conditions we discussed lead to a straightforward method for solving Problem~\ref{prob:finite}. We refer to this method as F-BCS which stands for finite BCS.
When $N=1$, F-BCS solves a CS problem for each $P\in\Psi$
\begin{equation}\label{eq:FBCSsingel_1}
    \hat{s}=\text{arg}\min_s{||s||_0} \text{ s.t. } b=APs,
\end{equation}
and chooses the sparsest $\hat{s}$. Under the uniqueness conditions it is the only one with no more than $k$ nonzero elements. Therefore if we know the sparsity level $k$ we can stop the search when we found a sparse enough $\hat{s}$. The recovered signal is $x=P\hat{s}$ where $P$ is the basis corresponding to the $\hat{s}$ we chose.
When $k$ is known an alternative method is to solve for each $P\in\Psi$
\begin{equation}\label{eq:FBCSsingel_2}
     \hat{s}=\text{arg}\min_s{||b-APs||^2_2} \text{ s.t. } ||s||_0<k,
\end{equation}
and choose $\hat{s}$ that minimizes $||b-AP\hat{s}||^2_2$. In the noiseless case this minimum is zero for the correct basis $P$.

When $N>1$ we can solve either \eqref{eq:FBCSsingel_1} or \eqref{eq:FBCSsingel_2} for each of the signals and select the sparsity basis according to the majority.

The solution to problems \eqref{eq:FBCSsingel_1} and \eqref{eq:FBCSsingel_2} can be approximated using one of the standard CS algorithms.
Since these algorithms are suboptimal, there is no guarantee that they provide the correct solution $x$, even for the correct basis $P$.
In general, when $k$ is small enough relative to $n$ these algorithms are known to perform very well. Moreover, when $N>1$, $P$ is selected according to the majority of signals, and therefore if the CS algorithm did not work well on a few of the signals it will not effect the recovery of the rest of the signals.

\subsection{F-BCS Simulation Results}
We now demonstrate the F-BCS method in simulation. We chose the set of bases $\Psi$ to contain 5 bases of size $64\times64$: the identity, DCT \cite{DCT}, Haar wavelet, Symlet wavelet and Biorthogonal wavelet \cite{wavelet}.
100 signals of length 64 were created randomly by generating random sparse vectors and multiplying them by the Biorthogonal wavelet basis in $\Psi$. Each sparse vector contained up to 6 nonzero elements in uniformly random locations, and values from a normal distribution.

The measurement matrix $A$ was an i.i.d Gaussian matrix of size $32\times 64$. The measurements were calculated first without noise, that is $B=AX$, and then with additive Gaussian noise with varying SNR from 30dB to 5dB.
For each noise level the F-BCS method was performed, where the CS algorithm we used was OMP \cite{OMP}.

Table~\ref{tab:F-BCS} summarizes the results. For all noise levels the basis selection according to the majority was correct. The miss detected column in the table contains the percentage of signals that indicated a false basis.  The average error column contains the average reconstruction error, calculated as the average of
\begin{equation}\label{eq:Err}
e_i=\frac{||x_i-\hat{x}_i||_2}{||x_i||_2}
\end{equation}
where $x_i,\hat{x}_i$ are the columns of the real signal matrix $X$ and the reconstructed signal matrix $\hat{X}$ respectively. The average is performed only on the signals that indicated the correct basis. The reconstruction of the rest of the signals obviously failed.
As can be seen from Table~\ref{tab:F-BCS} in the noiseless case the recovery is perfect and the error grows with the noise level. For high SNR there are no false reconstructions, but as the SNR decreases beyond 15dB the percentage of false reconstructions increases. In these cases, one should use more then one signal, such that if one of the signals failed there will be an indication for this through the rest of the signals.

\begin{table}[tbp]
\caption{f-bcs simulation results}\label{tab:F-BCS}
\begin{center}
\begin{tabular}{|c|c|c|}
\hline
SNR& Miss & Average  \\
& Detected & Error\\
\hline
$\infty$ & 0\% & $10^{-14}$\% \\
30dB & 0\% & 1.3\% \\
25dB & 0\% & 2.7\% \\
20dB & 0\%& 5.4\% \\
15dB & 1\%& 11.6\% \\
10dB & 12\% & 22.5\% \\
5dB & 25\% & 40.1\% \\
\hline
\end{tabular}
\end{center}
\end{table}

Another simulation we performed investigated the influence of the sparsity level $k$, which is the number of nonzero elements in $S$. The settings of this simulation were the same as those of the first simulation, only this time there was no noise added to the measurements, and $k$ was gradually increased from 1 to 32.
For each sparsity level new signals were generated with the same sparsity basis and measured by the same measurement matrix. For $k<8$ the recovery of the signal was perfect, but as expected, for higher values of $k$ the number of false reconstructed signals and the average error grew. The reason for this is that the OMP algorithm works well with small values of $k$, for higher values of $k$, even if the uniqueness conditions are still satisfied, the OMP algorithm may not find the correct solution.

\section{Sparse Basis}\label{sec:sparse}
A different constraint that can be added to Problem~\ref{prob:original} in order to reduce the number of solutions is the sparsity of the basis $P$. That is, we assume that the columns of the basis $P$ are sparse under some known dictionary $\Phi$, so that there exists some unknown sparse matrix $Z$ such that $P=\Phi Z$. We assume the number of nonzero elements in each column of $Z$ is known to equal $k_p$. We refer to $\Phi$ as a dictionary since it does not have to be square. Note that in order for $P$ to be a basis $\Phi$ must have full row rank, and $Z$ must have full column rank.

The constrained BCS in this case is then:
\begin{problem} \label{prob:sparse}
    Given the measurements $B$, the measurement matrix $A$ and the dictionary $\Phi$, which has full row rank, find the signal matrix $X$ such that $B=AX$ where $X=\Phi Z S$ for some $k$-sparse matrix $S$ and $k_p$-sparse and full column rank matrix $Z$.
\end{problem}

This problem is similar to that studied in \cite{sparseKSVD} in
the context of sparse DL. The difference is that \cite{sparseKSVD}
finds the matrices $Z,S$, while we are only interested in their
product. The motivation behind Problem~\ref{prob:sparse} is to
overcome the disadvantage of the previously discussed
Problem~\ref{prob:finite} in which the bases are fixed. When using
a sparse basis we can choose a dictionary $\Phi$ with fast
implementation, but enhance its adaptability to different signals
by allowing any sparse enough combination of the columns of
$\Phi$. Note that we can solve the problem separately for several
different dictionaries $\Phi$, and choose the best solution. This
way we can combine the sparse basis constraint and the constraint
of a finite set of bases. Another possible combination between
these two approaches is to define the basic dictionary as
$\Phi=[P_1,P_2,...]$, where the finite set of bases is
$\Psi=\{P_1,P_2,...\}$. This way we allow any sparse enough
combination of columns from all the bases in $\Psi$.

\subsection{Uniqueness Conditions}
As we now show, here too under appropriate conditions the constrained problem has a unique solution even when there is only one signal $N=1$. Therefore, instead of matrices $X,S,B$ we deal with vectors $x,s,b$ respectively.
Since $||s||_0\leq k$ and $Z$ is $k_p$-sparse, the vector $c=Zs$ necessarily satisfies $||c||_0\leq k_pk$. Therefore, Problem~\ref{prob:sparse} as
\begin{equation}\label{eq:Sparse_P01}
    \hat{c}=\text{arg}\min_c{||c||_0}\qquad \text{s.t. } b=A\Phi c,
\end{equation}
or equivalently:
\begin{equation}\label{eq:Sparse_P0}
    \hat{c}=\text{arg}\min_c{||b-A\Phi c||_2^2}\qquad \text{s.t. } ||c||_0\leq k_pk,
\end{equation}
where the recovery is $x=\Phi\hat{c}$.
The solutions to \eqref{eq:Sparse_P01} and \eqref{eq:Sparse_P0} are unique if $\sigma(A\Phi)\geq2k_pk$.
If there is more then one signal, $N>1$, then one can solve \eqref{eq:Sparse_P01} and \eqref{eq:Sparse_P0} for each signal separately.

Note that in Problem~\ref{prob:sparse} the matrix $Z$ necessarily has full column rank, while this constraint is dropped in \eqref{eq:Sparse_P01} and \eqref{eq:Sparse_P0}. However, if the solution without this constraint is unique then obviously the solution with this constraint is also unique.
Therefore, a sufficient condition for the uniqueness of Problem~\ref{prob:sparse} is $\sigma(A\Phi)\geq2k_pk$.

\subsection{Algorithms For Sparse BCS}
\subsubsection{Direct Method}
When there is only one signal, according to the uniqueness discussion, the solution to Problem~\ref{prob:sparse} can be found by solving either \eqref{eq:Sparse_P01} or \eqref{eq:Sparse_P0} using a standard CS algorithm.
When there are more signals the same process can be performed for each signal separately.
Since we use a standard CS algorithm, for this method to succeed we require the product $k_pk$ to be small relative to $n$.

\subsubsection{Sparse K-SVD}
The sparse K-SVD algorithm \cite{sparseKSVD} is a DL algorithm
that seeks a sparse dictionary. That is, given the measurements
$B$ and a base dictionary $D$ it finds $k_p$-sparse $Z$ and
$k$-sparse $S$, such that $B=D ZS$. In our case we can run sparse
K-SVD on $B$ with $D=A\Phi$ in order to find $Z$ and $S$, and then
recover the signals by $X=\Phi ZS$. The sparse K-SVD algorithm is
a variation of the K-SVD algorithm \cite{KSVD}, which is a popular
DL algorithm. Sparse K-SVD consists of two alternating steps. The
first is sparse coding, in which $Z$ is fixed and $S$ is updated
using a standard CS algorithm. The second step is dictionary
update, in which the support of $S$ is fixed and $Z$ is updated
together with the value of the nonzero elements in $S$. The
difference between sparse K-SVD and K-SVD is only in the
dictionary update step. Since the sparse K-SVD is a DL algorithm,
it requires a large number of diverse signals. Moreover, the
required diversity of the signals can prevent the algorithm from
working, for instance in cases of block sparsity.

In general, BCS cannot be solved using DL methods. However, under the sparse basis constraint BCS is
reduced to a problem that can be viewed as constrained DL, and therefore solved using sparse K-SVD.
Nevertheless, Problem~\ref{prob:sparse} is not exactly constrained DL, since in DL we seek the matrices $S$ and $Z$ themselves, whereas here we are interested only in their product $X=\Phi ZS$. Moreover, as in any DL algorithm, for sparse K-SVD to perform well it requires many diverse signals. However, for the uniqueness of Problem~\ref{prob:sparse} or for the direct method of solution, there is no need for such a requirement. The sparse K-SVD algorithm is also much more complicated than the direct method.

Nonetheless, sparse K-SVD has one advantage over the direct method in solving Problem~\ref{prob:sparse}. The direct method uses a standard CS algorithm in order to find $C=ZS$ which is $k_pk$-sparse. This algorithm provides the correct result only if the product $k_pk$ is small enough relative to $n$. On the other hand, the standard CS algorithms used in sparse K-SVD attempt to find separately $S$ which is $k$-sparse and $Z$ which is $k_p$-sparse, and therefore require $k$ and $k_p$ themselves to be small instead of the product $k_pk$.
Thus, when there are few signals, or even just one, and when $k_pk$ is small relative to $n$, then Problem~\ref{prob:sparse} should be solved using the direct method.
If $k_pk$ is large but still satisfies $\sigma(A\Phi)\geq2k_pk$, and if there are enough diverse signals, then  sparse K-SVD should be used.

\subsection{Simulation Results}
Simulation results for sparse K-SVD can be found in \cite{sparseKSVD}. Here we present simulation results for the direct method.
First of all we tested the influence of the sparsity level of the basis. We generated a random sparse matrix - $Z$, of size $256\times256$ with up to $k_p=6$ nonzero elements in each column. The value of $k$ - the number of nonzero elements in $S$, was gradually increased from 1 to 20. For each $k$ we generated $S$ as a random $k$-sparse matrix of size $256\times 100$, and created the signal matrix $X=\Phi ZS$, where $\Phi$ was the DCT basis.
$X$ was measured using a random Gaussian matrix $A$ of size $128\times256$, resulting in $B=AX$.

\begin{figure}[tbp]
    \begin{center}
     \includegraphics*[scale=0.5]{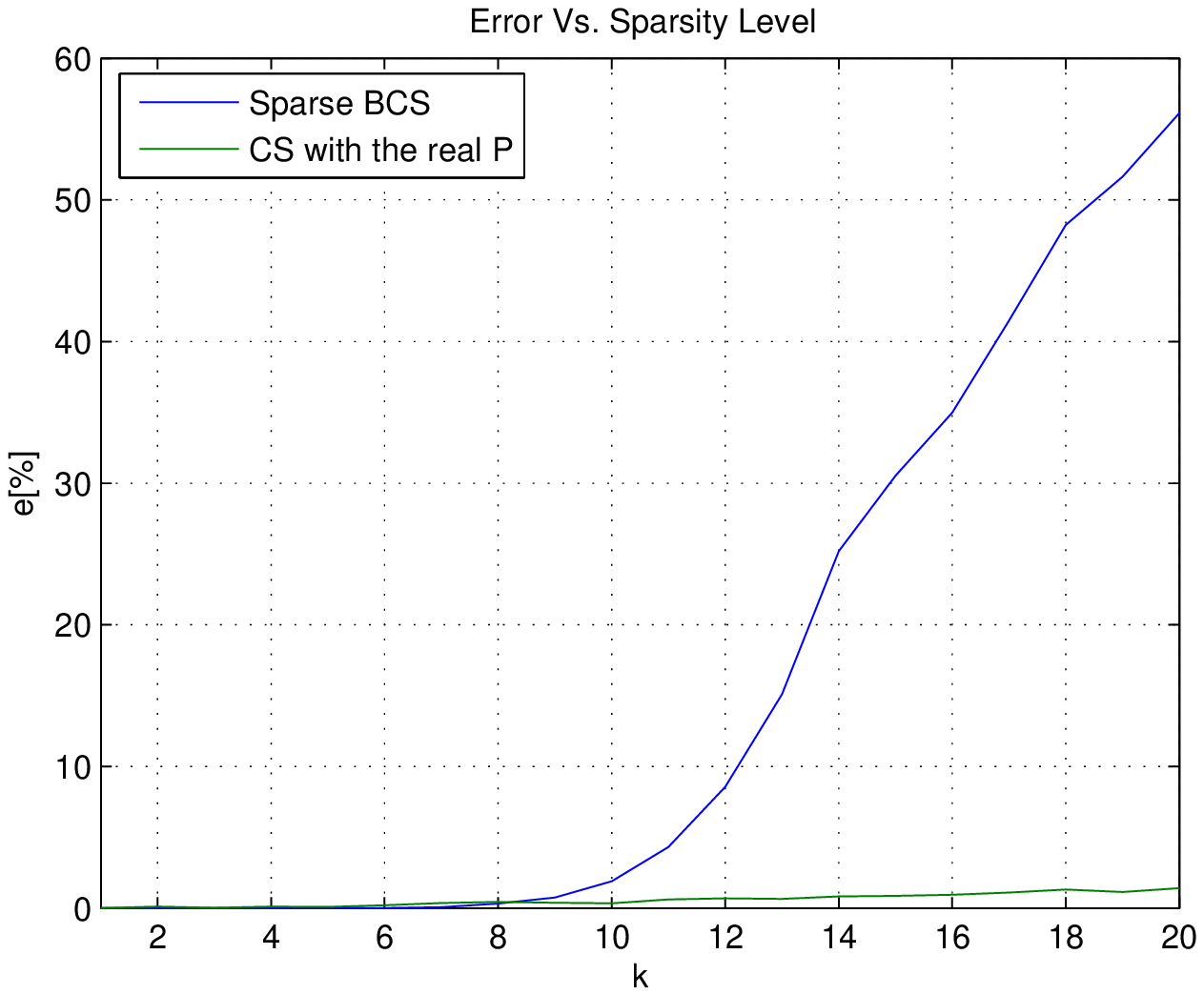}
     \caption{Reconstruction error as a function of the sparsity level} \label{fig:sparse}
    \end{center}
\end{figure}

We solved Problem~\ref{prob:sparse} given $A$ and $B$ using the direct method, where again the CS algorithm we used was OMP.
For comparison we also performed OMP with the real basis $P$, which is unknown in practice.
Fig~\ref{fig:sparse} summaries the results. For every value of $k$ the error of each of the graphs is an average over the reconstruction errors of all the signals, calculated as in \eqref{eq:Err}. Both the errors are similar for $k\leq 8$, but for larger $k$'s the error of the blind method is much higher.

Since $A$ is an i.i.d Gaussian matrix and the DCT matrix is orthogonal with probability~1, $\sigma(A\Phi)=129$. Therefore with probability~1 the uniqueness of the sparse BCS method is achieved as long as $k_pk\leq 64$, or $k\leq 10$. The error began to grow before this sparsity level because OMP is a suboptimal algorithm that is not guaranteed to find the solution even when it is unique, but works well on sparse enough signals.
The reconstruction error of the OMP which used the real $P$ grows much less for the same values of $k$. That is since in this case $k$ itself, instead of $k_pk$, should be small relative to $n$.

Sparse K-SVD can improve the results for high value of $k$, assuming of course it is small enough for the solution to be unique. However, in this simulation the number of signals is even less then the length of the vectors, and sparse K-SVD does not work well with such a small number of signals. In the sparse K-SVD simulations which are presented in \cite{sparseKSVD} the number of signals is at least 100 times the length of the signals.

We also investigated the influence of noise on the algorithm. The setting of this simulations were the same as in the previous simulation only this time we fixed $k=3$ and added Gaussian noise to the measurements $B$.
We looked at different noise levels, and for each level we ran the direct method for sparse BCS, and also for comparison we ran an OMP algorithm which used the real basis $P$.
Table~\ref{tab:sparse} summarizes the average errors of each of the methods. In the noiseless case there is a perfect recovery in both cases. As the SNR decreases both errors increases, but as can be expected, the one of the BCS grows faster. The reason for the big difference in the low SNR cases is again the fact that in the CS case the OMP algorithm is performed on sparser signals, relative to the sparse BCS case.

\begin{table}[tbp]
\caption{reconstruction error for different noise levels}\label{tab:sparse}
\begin{center}
\begin{tabular}{|l|c|r|}
\hline
SNR & CS & sparse BCS\\
\hline
$\infty$ & $10^{-14}$\% & $10^{-14}$\% \\
30dB & 1.2\% & 2.8\% \\
25dB & 1.5\% & 5.8\% \\
20dB & 3.3\% & 11.9\% \\
15dB & 7.1\% & 23.5\% \\
\hline
\end{tabular}
\end{center}
\end{table}

\section{Structural Constraint}\label{sec:struct}
The last constraint we discuss is a structural constraint on the basis $P$. We require $P$ to be block diagonal and  orthogonal.
The motivation for the block diagonal constraint comes form Problem~\ref{prob:new}, which looks for $\tilde{P}$ such that $\tilde{D}=A\tilde{P}$.
Assume for the moment that $\tilde{P}$ is block diagonal, such that:
\begin{equation*}
    \tilde{P}=\left[\begin{array}{ccc}\tilde{P}_1& & \\ & \ddots & \\ & & \tilde{P}_L \\ \end{array}\right],
\end{equation*}
and $A$ is chosen to be a union of orthonormal bases, as in \cite{Unions,Unions_tech,UncerPrinc,UncerPrincPair ,SparseRepUnion}. That is, $A=[A_1,...A_L]$ where $A_1,...,A_L$ are all orthonormal matrices. In this case
\begin{equation*}
D=[D_1,...,D_L]=[A_1P_1,...,A_LP_L],
\end{equation*}
and we can simply recover $\tilde{P}$ by:
\begin{equation}\label{eq:tildePrec}
    \tilde{P}=\left[\begin{array}{ccc}A_1^TD_1& & \\ & \ddots & \\ & & A_L^TD_L \\ \end{array}\right].
\end{equation}
Therefore, the solution to Problem~\ref{prob:new} under the constraint that $\tilde{P}$ is block diagonal is very simple.

Under the richness and spark conditions the BCS problem, as defined in Problem~\ref{prob:original}, is equivalent to Problem~\ref{prob:new}, where the basis $\tilde{P}$ in Problem~\ref{prob:new} is a column signed permutation of the basis $P$ in Problem~\ref{prob:original}. Since we are interested in the solution to Problem~\ref{prob:original}, the constraint should be on the basis $P$ instead of $\tilde{P}$.
However, if we constrain $P$ to be block diagonal, then the solution to the equivalent Problem~\ref{prob:new} is not as simple as in \eqref{eq:tildePrec}. In Problem~\ref{prob:new} we look for $\tilde{P}=PQ$, for some unknown signed permutation matrix $Q$. Under the block diagonal constraint on $P$ the matrix $\tilde{P}=PQ$ is not necessarily block diagonal, and therefore we cannot use  \eqref{eq:tildePrec} to recover it.

We can guarantee that $\tilde{P}$ is block diagonal only if we can guarantee that $Q$ is block diagonal. That is, $Q$ permutes only the columns inside each block of $P$, and does not mix the blocks or change the outer order of them.
As we prove below in the uniqueness discussion, this can be guaranteed if we require $P$ to have more blocks than $A$. Specifically, we require $P$ to have $2L$ blocks, which is twice the number of blocks in $A$. Such a basis $P$ is called \emph{$2L$-block diagonal}.
In fact, the number of blocks in $P$ can be $ML$ for any integer $M\geq2$. We use $M=2$ for simplicity; the expansion to $M>2$ is trivial.

We also constraint $P$ to be orthogonal. The motivation for this is the spark condition. In order be able to solve Problem~\ref{prob:new} instead of Problem~\ref{prob:original}, we need to satisfy $\sigma(AP)\geq2k$. By constraining $P$ to be orthogonal we can use results similar to Proposition~\ref{prop:spark} in order to achieve this requirement with probability~1.

The constrained BCS problem is then:
\begin{problem}\label{prob:struct}
    Given the measurements $B$ and the measurement matrix $A\in\mathbb{R}^{n\times nL}$ find the signal matrix $X$
    such that $B=AX$ where $X=PS$ for some orthogonal $2L$-block diagonal matrix $P$ and $k$-sparse matrix $S$.
\end{problem}

In this new settings the size of the measurement matrix $A$ is $n \times nL$, where $n$ is the number of measurements and $L$ is the number of $n\times n$ blocks in $A$, which equals the compression ratio. Moreover, The length of the signals is $m=nL$, and the size of the basis $P$ is $nL\times nL$. Since $P$ is $2L$-block diagonal, the size of its blocks is $\frac{n}{2}\times\frac{n}{2}$. Therefore, $n$ must be even.

This constrained problem can be useful for instance in multichannel systems, where the signals from each channel are sparse under separate bases. In such systems we can construct $X$ by concatenating signals from several different channels, and compressively sampling them. For example, in microphone arrays \cite{MicArray} or antenna arrays \cite{AntennaArray}, we can divide the samples from each microphone / antenna into time intervals in order to obtain the ensemble of sampled signals $B$. Each column of $B$ is a concatenation of the signals from all the microphones / antennas over the same time interval.

\subsection{Uniqueness Conditions}
To ensure a unique solution to Problem~\ref{prob:struct}, we need the DL solution given $B$ to be unique. Therefore, we assume that the richness conditions on $S$ and the spark condition on $AP$ are satisfied. Then, Problem~\ref{prob:struct} is equivalent to the following problem:
\begin{problem}\label{prob:structEqui}
    Given the matrices $\tilde{D}$ and $A$, which have more columns then rows, find an orthogonal $\tilde{P}$ such that $\tilde{D}=A\tilde{P}$, and $\tilde{P}=PQ$ for some signed permutation matrix $Q$ and orthogonal $2L$-block diagonal matrix $P$.
\end{problem}
In order to discuss conditions for uniqueness of the solution to Problem~\ref{prob:structEqui} we introduce the following definition.

\begin{definition}\label{def:InterBlock}
    Denote $A=[A_1,...,A_L]$, such that $A_i\in\mathbb{R}^{n\times n}$ for any $1\leq i\leq L$.
    $A$ is called \emph{inter-block diagonal} if there are two indices $i\neq j$ for which the product:
    \begin{equation*}
    A_i^TA_j=\left[\begin{array}{cc}R_1&R_2\\R_3&R_4\end{array}\right],
    \end{equation*}
    satisfies:
    \begin{equation*}
    \begin{split}
    \lefteqn{\text{rank}(R_1)=\text{rank}(R_4)}\\
    & \text{rank}(R_2)=\text{rank}(R_3)=\frac{n}{2}-\text{rank}(R_1).
    \end{split}
    \end{equation*}
    In particular if the product $A_i^TA_j$ is 2-block diagonal then $A$ is inter-block diagonal.
\end{definition}

With this definition in hand we can now define the conditions for the uniqueness of Problem~\ref{prob:structEqui}.

\begin{theorem}\label{th:struct}
    If $A\in\mathbb{R}^{n\times nL}$ is a union of $L$ orthogonal bases, which is not inter-block diagonal, and $\sigma(AP)=n+1$, then the solution to Problem~\ref{prob:structEqui} is unique.
\end{theorem}

The proof of this theorem uses the next lemma.
\begin{lemma}\label{lemma:PQ}
    Assume $P$ and $\hat{P}$ are both orthogonal $2L$-block diagonal matrices, and $A$ satisfies the conditions of Theorem~\ref{th:struct}. If $A\hat{P}=APQ$ for some signed permutation matrix $Q$, then $\hat{P}=PQ$.
\end{lemma}

In general since $A$ has a null space, if the matrices $A,P,\hat{P}$ did not have their special structures, then the equality $A\hat{P}=APQ$ would not imply $\hat{P}=PQ$. However, according to Lemma~\ref{lemma:PQ} under the constraints on $A,P,\hat{P}$ this is guaranteed.
The full proof of Lemma~\ref{lemma:PQ} appears in Appendix A. Here we present only the proof sketch.

\emph{Proof sketch}:
It is easy to see that due to the orthogonality of the blocks of $A$, if $Q$ is block diagonal then $A\hat{P}=APQ$ implies $\hat{P}=PQ$. Therefore, we need to prove that $Q$ is necessarily block diagonal.
Denote $D=AP$. In general the multiplication $DQ$ can yield three types of changes in $D$. It can mix the blocks of $D$, permute the order of the blocks of $D$, and permute the columns inside each block. $Q$ is block diagonal if and only if it permutes only the columns inside each block, but does not mix the blocks or change their outer order.
First we prove that $Q$ cannot mix the blocks of $D$. For this we use the condition on the spark of $D$, and the orthogonality of the blocks.
Next we prove that $Q$ cannot change the outer order of the blocks. This time we use the fact that both $P$ and $\hat{P}$ have $2L$ blocks and that $A$ is not inter-block diagonal.
Therefore, $Q$ can only permute the columns inside each block, which implies it is block diagonal
\Qed

If $P$ and $\tilde{P}$ have only $L$ blocks instead of $2L$, then $Q$ can change the outer order of the blocks of $D$, such that it does not have to be block diagonal. Therefore, if the constraint on $P$ was that it has $L$ blocks instead of $2L$, then Lemma~\ref{lemma:PQ} would be incorrect, such that the solution to the Problem~\ref{prob:structEqui}, and therefore to Problem~\ref{prob:struct}, would not be unique. On the other hand the extension of the proof of Lemma~\ref{lemma:PQ} to $ML$ blocks where $M>2$ is trivial.

\emph{Proof of Theorem~\ref{th:struct}}:
The proof we provide for Theorem~\ref{th:struct} is constructive, although far from being a practical method to deploy in practice. Denote the desired solution of Problem~\ref{prob:structEqui} by $\tilde{P}=PQ$, and denote:
\begin{equation*}
    A=[A_1,...,A_L] \; , \; P=\left[\begin{array}{ccc} P^1& &\\& \ddots & \\ & & P^{2L} \end{array}\right],
\end{equation*}
where $A_i$ for $i=1,..,L$ and $P^j$ for $j=1,...,2L$ are all orthogonal matrices.

We first find a permutation matrix $Q_D$ such that $\hat{D}=\tilde{D}Q_D=A\hat{P}$, where $\hat{P}$ is an orthogonal $2L$-block diagonal matrix. There is always at least one such permutation. For instance, we can choose $Q_D$ to equal the absolute value of $Q^T$. In this case $\hat{P}$ equals $P$ up to the signs, and therefore it is necessarily orthogonal $2L$-block diagonal.

Denote the blocks of $\hat{P}$ by $\hat{P}^j$ for $j=1,...,2L$, and note that
\begin{eqnarray*}
    \lefteqn{\hat{D}=[\hat{D}_1,...,\hat{D}_L]=}\\
    & \Big[ A_1\left(\begin{array}{ccc} \hat{P}^1& \\ &
    \hat{P}^2 \end{array}\right),\ldots,A_L\left(\begin{array}{ccc} \hat{P}^{2L-1}& \\ & \hat{P}^{2L}
    \end{array}\right)\Big].
\end{eqnarray*}
Since $A_i$ are orthogonal for all $i=1,...,L$, we can recover the blocks of $\hat{P}$ by
\begin{equation*}
    \left[\begin{array}{ccc} \hat{P}^{2i-1} & \\ & \hat{P}^{2i} \end{array}\right]=A_i^T\hat{D}_i,
\end{equation*}
such that
\begin{equation*}
    \hat{P}=\left[\begin{array}{ccc} A_1^T\hat{D}_1& & \\ & \ddots & \\ & & A_L^T\hat{D}_L \end{array}\right].
\end{equation*}
Since both $P$ and $\hat{P}$ are orthogonal $2L$-block diagonal, according to Lemma~\ref{lemma:PQ} the equality $\hat{D}=A\hat{P}=APQQ_D$ implies $\hat{P}=PQQ_D$. Therefore, we can recover $\tilde{P}$ by $\tilde{P}=PQ=\hat{P}Q_D^T$.
\Qed

The conclusion from Theorem~\ref{th:struct} is that if the richness conditions on $S$ are satisfied and $A$ satisfies the conditions of Theorem~\ref{th:struct}, then the solution to Problem~\ref{prob:struct} is unique.

As proven in Appendix B one way to guarantee that $A$ satisfies the conditions of Theorem~\ref{th:struct} with probability~1 is to generate it randomly from an i.i.d Gaussian distribution and perform a Gram Schmidt process on each block in order to make it orthogonal. This claim is similar to Proposition~\ref{prop:spark} except that the statistics of $A$ is a bit different due to the Gram Schmidt process.

\subsection{The OBD-BCS Algorithm}\label{sec:OBDalg}
Although the uniqueness proof is constructive it is far from being practical.
In order to solve Problem~\ref{prob:struct} by following the uniqueness proof one needs to perform a DL algorithm on $B$, resulting in $\tilde{D},\tilde{S}$. Then go over all the permutations $\hat{D}=\tilde{D}Q_D$, and look for $Q_D$ such that the matrices $A_i^T\hat{D}_i$, for all $i=1,...,L$, are 2-block diagonal. After finding such a permutation the recovery of $X$ is
\begin{equation*}
    X=\left[\begin{array}{ccc} A_1^T\hat{D}_1& & \\ & \ddots & \\ & & A_L^T\hat{D}_L \end{array}\right]Q_D^T\tilde{S}.
\end{equation*}

The problem with this method is the search for the permutation $Q_D$. There are $m!$ different permutations of the columns of $D$, where $m=nL$ is the length of the signals, while only $[(\frac{m}{2L})!]^{2L}$ of them satisfy the requirement (see Appendix C). As $m$ and $L$ grow the relative fraction of the desirable permutations decreases. For instance, for signals of length $m=16$ and a compression ratio of $L=2$ only $1.58\cdot10^{-6}\%$ of the permutations satisfy the requirement. For the same signals but a higher compression ratio of $L=4$ only $1.22\cdot10^{-9}\%$ satisfy the condition, and for longer signals of length $m=64$ and $L=2$ only $1.51\cdot10^{-34}\%$ satisfy the requirement.

Therefore, a systematic search is not practical, even for short signals. Moreover, in practice the output of the DL algorithm contains some error, so that even for the correct permutation the matrices $A_i^{-1}\hat{D}_i$ are not exactly 2-block diagonal, which renders the search even more complicated. Although there exist suboptimal methods for permutation problems such as \cite{HGA}, these techniques are still computationally extensive and are sensitive to noise.

Instead we present the orthogonal block diagonal BCS (OBD-BCS) algorithm for the solution of Problem~\ref{prob:struct}, which is, in theory, equivalent to DL followed by the above postprocessing. However, it is much more practical and simple.
This algorithm is a variation of the DL algorithm in \cite{Unions, Unions_tech}, which learns a dictionary under the constraint that the dictionary is a union of orthogonal bases. Given $B$ the algorithm in \cite{Unions, Unions_tech} aims to solve
\begin{eqnarray}\label{eq:UnionsEq}
    \lefteqn{\quad \min_{D,S}{||B-DS||_F^2}}\\
    & \text {s.t. $S$ is $k$-sparse and $D$ is a union of orthogonal bases.} \nonumber
\end{eqnarray}
In the BCS case $P$ is orthogonal $2L$-block diagonal and $A$ is a union of $L$ orthogonal bases. Therefore, the equivalent dictionary is:
\begin{eqnarray*}
    \lefteqn{D=AP=} \\
    & \Big[ A_1\left(\begin{array}{ccc} P^1& \\ & P^2 \end{array}\right), \ldots
    ,A_L\left(\begin{array}{ccc} P^{2L-1}& \\ & P^{2L} \end{array}\right)\Big].
\end{eqnarray*}
Since all $A_i$ and $P^i$ are orthogonal, here too $D$ is a union of orthogonal bases. The measurement matrix $A$ is known and we are looking for an orthogonal $2L$-block diagonal matrix $P$ and a sparse matrix $S$ such that $B=APS$. This leads to the following variant of \eqref{eq:UnionsEq}:
\begin{eqnarray}\label{eq:OBDobj}
    \lefteqn{\quad \min_{P,S}{||B-APS||_F^2}}\\
    & \text{s.t. $S$ is $k$-sparse and $P$ is orthogonal $2L$-block diagonal.}\nonumber
\end{eqnarray}

The algorithm in \cite{Unions, Unions_tech} consists of two alternating steps. The first step is sparse coding, in which the dictionary $D$ is fixed and the sparse matrix $S$ is updated. The second step is dictionary update, in which $S$ is fixed and $D$ is updated. This algorithm finds the dictionary $D=AP$ and the sparse matrix $S$ but not the basis $P$, and consequently, not the signal matrix $X=PS$.

In OBD-BCS we follow similar steps. The first step is again sparse coding, in which $P$ is fixed and $S$ is updated. The second step is basis update, in which $S$ is fixed and $P$ is updated. The difference between OBD-BCS and the algorithm in \cite{Unions, Unions_tech} is mainly in the second step, where we add the prior knowledge of the measurement matrix $A$ and the block diagonal structure of $P$. In addition, we use a different CS algorithm in the sparse coding step.

We now discuss in detail the two steps of OBD-BCS.

\subsubsection{Sparse Coding}
In this step $P$ is fixed so that the optimization in \eqref{eq:OBDobj} becomes:
\begin{equation}\label{eq:Step1}
    \min_{S}{||B-APS||_F^2}\qquad \text{s.t. $S$ is $k$-sparse.}
\end{equation}
It is easy to see that \eqref{eq:Step1} is separable in the columns of $S$. Therefore, for each column of $B$ and $S$ we need to solve
\begin{equation}\label{eq:Step1sep}
    \min_{s}{||b-APs||_2^2}\qquad \text{s.t. } ||s||_0\leq k,
\end{equation}
where $s,b$ are the appropriate columns of $S,B$ respectively. This is a standard CS problem, as in \eqref{P0}, with the additional property that the combined measurement matrix $D=AP$ is a union of orthogonal bases. This property is used by the block coordinate relaxation (BCR) algorithm \cite{Unions, Unions_tech, BCR}. The idea behind this algorithm is to divide the elements of $s$ into blocks corresponding to the orthogonal blocks of $D$. In each iteration all the blocks of $s$ are fixed except one, which is updated using soft thresholding.
The DL algorithm proposed by \cite{Unions, Unions_tech} is a variation of the BCR algorithm, which aims to improve its convergence rate.
In OBD-BCS we can also use this variation. However, experiments showed that the results are about the same as the results with OMP. Therefore, we use OMP in order to update the sparse matrix $S$, when the basis $P$ is fixed.

\subsubsection{Basis Update}
In this step the sparse matrix $S$ is fixed and $P$ is updated.
Divide each of the $nL\times N$ matrices $S$ and $X$ into $2L$ submatrices of size $\frac{n}{2}\times N$ such that:
\begin{equation*}
    S=\left[\begin{array}{ccc} S^1 \\ \vdots  \\ S^{2L} \end{array} \right] \; , \;
    X=\left[\begin{array}{ccc} X^1  \\ \vdots  \\ X^{2L} \end{array} \right].
\end{equation*}
Divide each orthogonal block of $A$ into two blocks: $A_i=[A^{2i-1},A^{2i}]$ for $i=1,...,L$, such that:
\begin{equation*}
    A=[A_1,...,A_L]=[A^1,A^2,...,A^{2L-1},A^{2L}].
\end{equation*}
With this notation $X^i=P^iS^i$, and $B=\sum_{i=1}^{2L}{A^iP^iS^i}$. Therefore, \eqref{eq:OBDobj} becomes:
\begin{eqnarray}\label{eq:Step2full}
    \lefteqn{\min_{P^1,...,P^{2L}}{||B-\sum_{j=1}^{2L}{A^jP^jS^j}||_F^2}}\\
    & \text{s.t. $P^1,...,P^{2L}$ are orthogonal.} \nonumber
\end{eqnarray}
To minimize \eqref{eq:Step2full}, we iteratively fix all the blocks $P^j$ for $j=1,...,2L$ except one, denoted by $P^i$, and solve
\begin{equation}\label{eq:Step2}
    \min_{P^i}{||B^i-A^iP^iS^i||_F^2} \qquad \text{s.t. $P^i$ is orthogonal}
\end{equation}
where $B^i=B-\sum_{j\neq i}{A^jP^jS^j}$. With slight abuse of notation, from now on we abandon the index $i$.

Since $P$ is orthogonal and $A$ is constructed of columns from an orthogonal matrix, $P^TA^TAP=I$, and $||APS||_F^2=||S||_F^2$. Thus, \eqref{eq:Step2} reduces to
\begin{equation}\label{eq:Step2mod}
    \max_{P}\{\text{Tr }[B^TAPS]\} \qquad \text{s.t. $P$ is orthogonal.}
\end{equation}
Let the singular value decomposition (SVD) of the matrix $R=SB^TA$ be $R=U\Sigma V^T$, where $U$, $V$ are orthogonal matrices and $\Sigma$ is a diagonal matrix. Using this notation we can manipulate the trace in \eqref{eq:Step2mod} as follows:
\begin{equation*}
    \text{Tr}[B^TAPS]=\text{Tr}[SB^TAP]=\text{Tr}[\Sigma V^TPU].
\end{equation*}
The matrix $Z=V^TPU$ is orthogonal if and only if $P$ is orthogonal. Therefore, \eqref{eq:Step2mod} is equivalent to
\begin{equation*}
    \max_{Z}\{\text{Tr }[\Sigma Z]\} \qquad \text{s.t. $Z$ is orthogonal.}
\end{equation*}
If the matrix $R=SB^TA$ has full rank then $\Sigma$ is invertible. In this case the maximization is achieved only for $Z=I$, and therefore $P^i=VU^T$ is the unique minimum of \eqref{eq:Step2}. Even if $R$ does not have full rank $P^i=VU^T$ achieves a minimum of \eqref{eq:Step2}.

\begin{table}[tbp]
\caption{the obd-bcs algorithm}\label{tab:OBD}
\begin{center}
\begin{tabular}{|l|c|r|}
\hline
\textbf{Inputs}:\\
    $\bullet$ $B\in\mathbb{R}^{n \times N}$ - measurements \\
    $\bullet$ $A\in\mathbb{R}^{n\times nL}$ - measurement matrix (union of $L$ orthogonal bases) \\
\hline
\textbf{Outputs}:\\
    $\bullet$ $\hat{X}\in\mathbb{R}^{nL \times N}$ - reconstructed signal matrix \\
\hline
\textbf{Algorithm}:\\
    $\bullet$ Initiate $\hat{P}=I$ (the identity).\\
    $\bullet$ Repeat until a stoping criteria is reached:\\
        $\qquad \circ$ \emph{Sparse coding}: find the sparsest $\hat{S}$ such that $B=A\hat{P}\hat{S}$,\\
        $\qquad \;\;$ for instance using OMP.\\
        $\qquad \circ$ \emph{Basis update}: for all $i=1,...,2L$:\\
        $\qquad \qquad$ Calculate $B^i=B-\sum_{j\neq i}{A^j\hat{P}^j\hat{S}^j}$.\\
        $\qquad \qquad$ Use SVD: $\hat{S}^i(B^i)^TA^i=U\Sigma V^T$.\\
        $\qquad \qquad$ Update: $\hat{P}^i=VU^T$.\\
    $\bullet$ Calculate: $\hat{X}=\hat{P}\hat{S}$.\\
\hline
\end{tabular}
\end{center}
\end{table}

Table~\ref{tab:OBD} summarize the OBD-BCS algorithm. Note that the initiation can be any $2L$-block diagonal matrix, not necessarily the identity matrix as written in the table; however, the identity matrix is simple to implement.
This algorithm is much simpler then following the uniqueness proof, which requires a combinatorial permutation search. Each iteration of the OBD-BCS algorithm uses a standard CS algorithm and $2L$ SVDs.

An important question that arises is whether the OBD-BCS algorithm converges. To answer this question we look at each step separately.
If the sparse coding step is performed perfectly it solves \eqref{eq:Step1} for the current $P$. That is, the objective of \eqref{eq:OBDobj} is reduced or at least stays the same. In practice, for small enough $k$ the CS algorithm converges to the solution of \eqref{eq:Step1}.
However, in order to guarantee the objective of \eqref{eq:OBDobj} is reduced or at least not increased in this step, we can always compare the new solution after this step with the one from the previous iteration and chose the best of them.

Note that this step is performed separately on each column of $S$. That is, we can choose to keep only some of the columns from the previous iteration, while the rest are updated. If at least part of the columns are updated then the next basis update step changes the basis $P$, so that in the following sparse coding step we can get a whole new matrix $S$.
Therefore, the decision to keep the results from the previous iteration does not imply we keep getting the same results in all the next iterations.
Another possibility is to keep only the support of the previous solution and update the values of the nonzero elements using least-squares.
In practice, in our simulations the algorithm converges even without any comparison to the previous iteration.

The basis update step is divided into $2L$ steps. In each, all the blocks of $P$ are fixed except one, which is updated to minimize \eqref{eq:Step2}. Therefore, the objective of \eqref{eq:Step2} is reduced or at least stays the same in each of the $2L$ steps constructing the basis update step. Therefore, the objective of \eqref{eq:Step2full}, which is equivalent to \eqref{eq:OBDobj} with fixed $S$, is reduced or not increased in the basis update step.

Thus, as in \cite{Unions, Unions_tech}, the algorithm we are based on, and as in other DL algorithms such as \cite{MOD, KSVD}, we cannot prove the OBD-BCS algorithm converges to the unique minimum of \eqref{eq:OBDobj}. However, we can guarantee that under specific conditions there is a unique minimum and that the objective function is reduced or at least stays the same in each step of the algorithm. Furthermore, as can be seen in the next section the OBD-BCS algorithm performs very well in simulations on synthetic data.

\subsection{OBD-BCS Simulations}\label{sec:OBDsim}
As in the first two constraints we evaluated the algorithm performance on synthetic data. The signal matrix $X$ had 64 rows and was generated as a product of a random sparse matrix - $S$ and a random orthogonal 4-block diagonal matrix - $P$. The value of the nonzero elements in $S$ was generated randomly from a normal distribution, and the four orthogonal blocks of $P$ were generated from a normal distribution followed by a Gram Schmidt process. The measurement matrix $A$ was constructed of two random $32\times 32$ orthogonal matrices, that were generated from a normal distribution followed by a Gram Schmidt process. The number of signals and the sparsity level were gradually changed in order to investigate their influence.

The stopping rule of the algorithm was based on a maximal number of iterations and the amount of change in the matrices $S$ and $P$. If the change from the last iteration was too small, or if the maximal number of iterations was reached, then the algorithm stopped. In most cases the algorithm stopped due to small change between iterations after about 30 iterations.

\begin{figure}[tbp]
    \begin{center}
     \includegraphics*[scale=0.6]{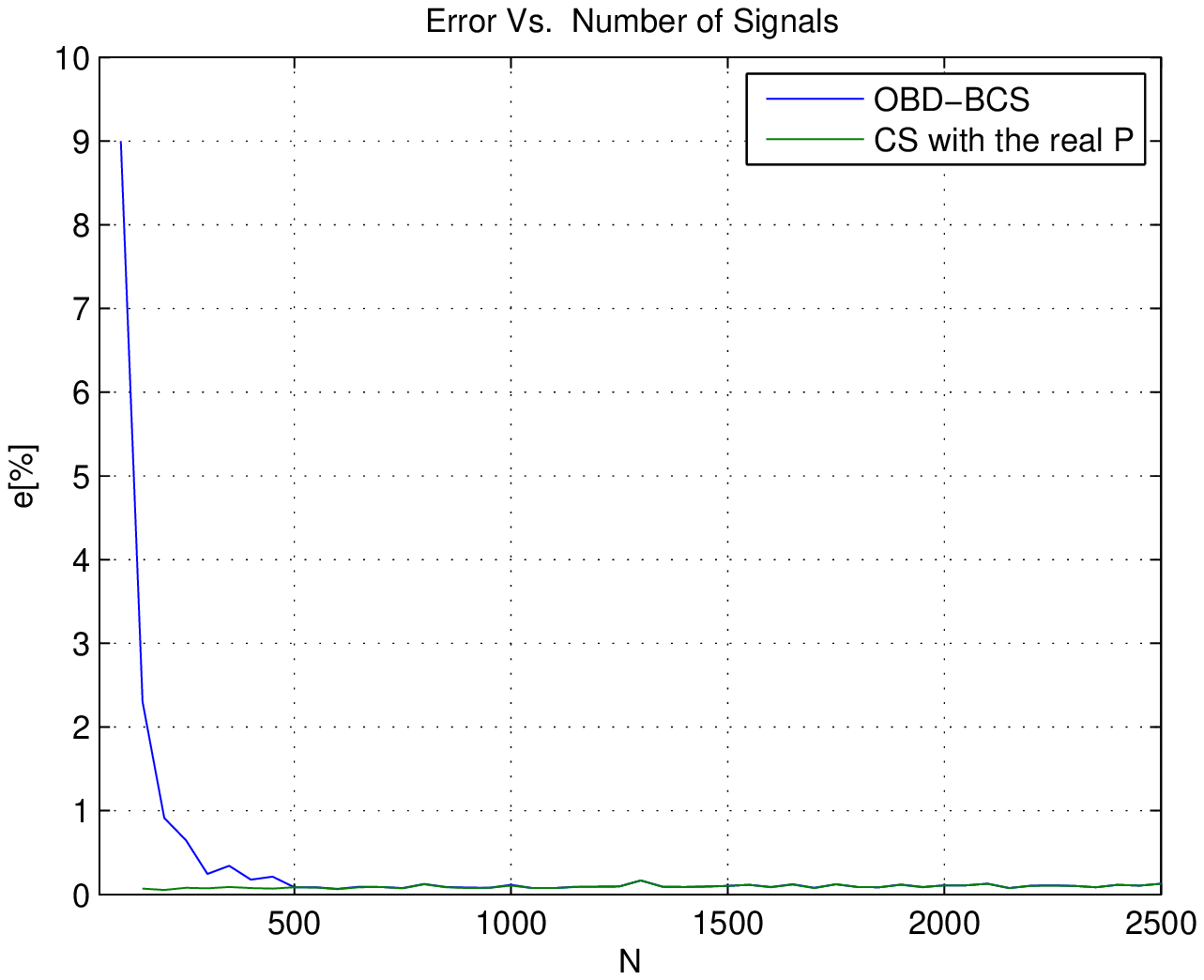}
     \caption{Reconstruction error as a function of the number of signals, for sparsity level of $k=4$.} \label{fig:e_Vs_n}
    \end{center}
\end{figure}

First we examined the influence of two parameters, $N$ - the number of signals needed for the reconstruction, and $k$ - the sparsity level. Fig.~\ref{fig:e_Vs_n} considers the influence of $N$ where the sparsity level is set to $k=4$. For each value of $N$ from 150 to 2500 the error presented in the upper graph is an average over 20 simulations of the OBD-BCS algorithm. In each simulation the sparse vectors and the orthogonal matrix where generated independently, but the measurement matrix was not changed. The error of each signal was calculated according to \eqref{eq:Err}.

For comparison, the lower graph in Fig.~\ref{fig:e_Vs_n} is the average error of a standard CS algorithm that was performed on the same data, and used the real basis $P$, which is unknown in practice. The CS algorithm we used was again OMP.
As expected, the results of the CS algorithm are independent of the number of signals, since it is performed separately and independently on each signal. The average error of this algorithm is 0.08\%. The reason for this nonzero error, although $P$ is known, is that for a small portion of the signals the OMP algorithm fails.

It is clear from Fig.~\ref{fig:e_Vs_n} that for $N>500$ the reconstruction results of the proposed algorithm are successful and similar to those obtained when $P$ is known.
Similarly to the conclusion in \cite{miki}, the reconstruction is successful even for $n$ much smaller then the number needed in order to satisfy the sufficient richness conditions, which is $\binom{m}{k}(k+1)\approx 3\cdot10^6$.
As in most DL algorithms, the algorithm in \cite{Unions, Unions_tech} was evaluated by counting the number of columns of the dictionary that are detected correctly. The conclusions of  \cite{Unions, Unions_tech} are that their algorithm can find about 80\% of the columns when the number of signals is at least $20n=640$, and can find all the columns when the number of signals is at least $50n=1600$. Using the same measurement matrix dimensions as in \cite{Unions, Unions_tech}, the minimal number of signals the OBD-BCS algorithm requires is only 500.

\begin{figure}[tbp]
    \begin{center}
     \includegraphics*[scale=0.6]{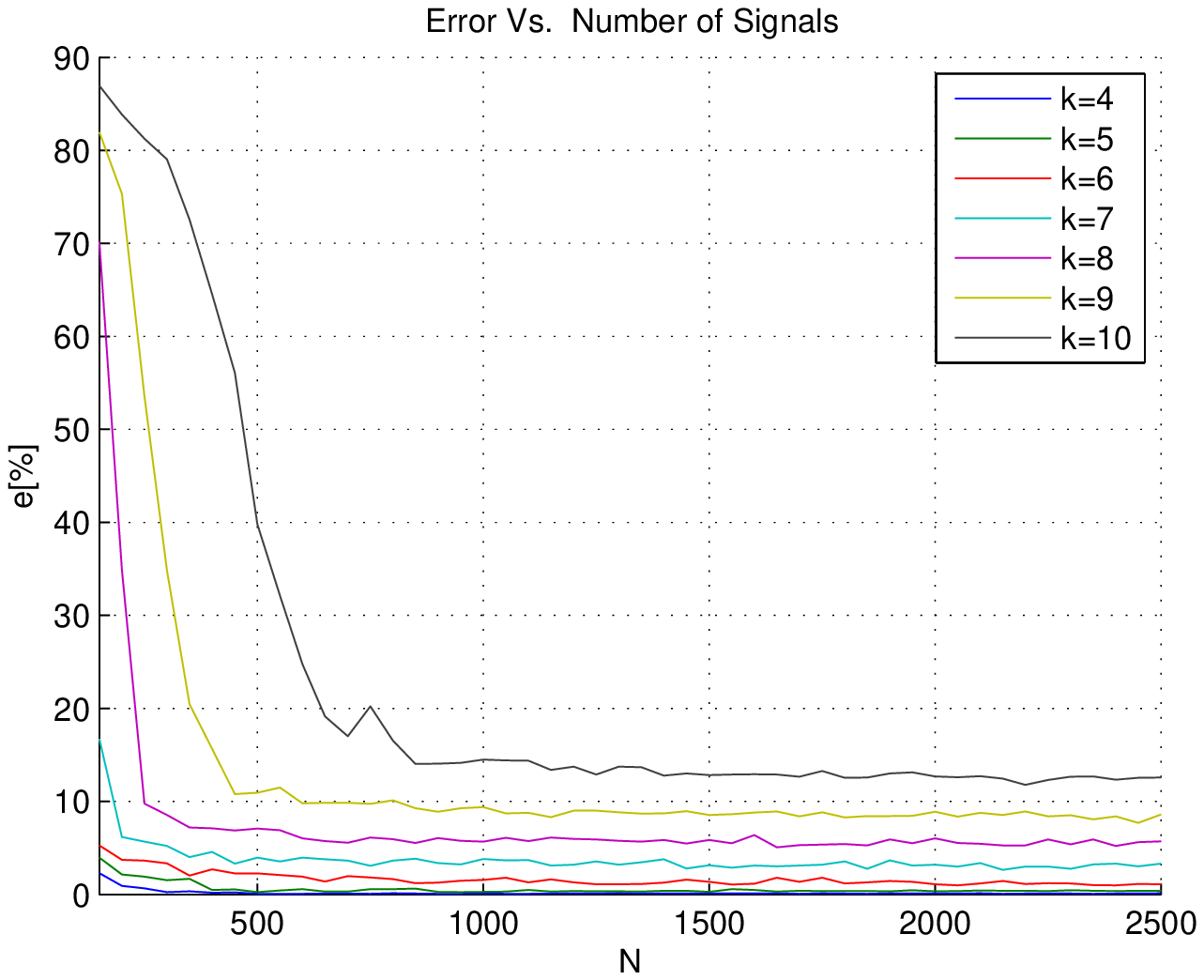}
     \caption{Reconstruction error as a function of the number of signals for different values of $k$.}\label{fig:e_Vs_nL}
    \end{center}
\end{figure}

In order to examine the influence of $k$ we performed the same experiment as before but for different values of $k\leq10$. The results are presented in Fig.~\ref{fig:e_Vs_nL}. It can be seen that for all values of $k$ the graph has the same basic shape: the error decreases with $N$ until a critical $N$, after which the error is almost constant. As $k$ grows this critical $N$ increases and so does the value of the constant error. The graphs for $k=1$, $k=2$, $k=3$ follow the same pattern; they are not in the figure since they are not visible on the same scale as the rest.

Next we investigated the influence of noise on the algorithm. In this simulation the noisy measurements $B$ were calculated as $B=APS+W$, where the elements of $W$ were white Gaussian noise. For each noise level 20 simulations were performed and the average error was calculated. In all simulations $k=4$ and $N=800$. Table~\ref{tab:noise} summarizes the results of the OBD-BCS algorithm and those of OMP algorithm which uses the real $P$.
It is clear from the table that in the noiseless case the error of both algorithms is similar, therefore in this case the prior knowledge of the basis $P$ can be avoided. As the SNR decreases both error increase, but the error of OBD-BCS algorithm increases a bit faster then that of the CS algorithm. However, the difference is not very big.

\begin{table}[tbp]
\caption{reconstruction error for different noise levels}\label{tab:noise}
\begin{center}
\begin{tabular}{|l|c|r|}
\hline
SNR & CS & OBD-BCS\\
\hline
$\infty$ &  0.008\% &  0.008\% \\
35dB & 0.82\% & 0.88\% \\
30dB & 1.54\% & 1.64\% \\
25dB & 2.95\% & 3.23\% \\
20dB & 5.81\% & 6.10\% \\
15dB & 12.03\% &12.58\% \\
10dB & 25.11\% &26.04\% \\
\hline
\end{tabular}
\end{center}
\end{table}

\section{Comparative Simulation}\label{sec:SimComp}
The following simulation illustrates the difference between the three BCS methods presented in this work.
In this simulation the length of the signals was $m=128$, the sparsity level was $k=6$, the number of signals was $N=2000$, and the compression ratio was $L=2$. The syntectic data was generated as in Section~\ref{sec:OBDsim}, but this time the instead of generating $P\in\mathbb{R}^{128\times128}$ randomly we used
\begin{equation*}
    P=\frac{1}{\sqrt{2}}\left[\begin{array}{ccc} \begin{array}{ccc} 1 & -1 \\ 1 & 1 \end{array}   & & \\
    & \ddots & \\
    & & \begin{array}{ccc} 1 & -1 \\ 1 & 1 \end{array} \end{array} \right],
\end{equation*}
which can be viewed as an orthogonal 4-block diagonal matrix (each block is 16-block diagonal by itself).

We used five different methods for the reconstruction of these signals.
\begin{enumerate}
\item CS algorithm with the real basis $P$.
\item CS algorithm with an estimated basis $P_{DL}$.
\item The F-BCS method.
\item The direct method for sparse BCS.
\item The OBD-BCS algorithm.
\end{enumerate}

In all the methods above we used OMP as the standard CS algorithm. The first method, came as a reference for the rest. It used the real basis $P$, whose knowledge we are trying to avoid.
The second method is an intuitive way to reconstruct the signals. Since the basis $P$ is unknown one can estimate it first and then perform a CS algorithm which uses the pre-estimated basis.
We performed the estimation using a training set of 2000 signals and a DL algorithm. The estimated basis is denoted by $P_{DL}$. There are several different DL algorithms, eg. \cite{MOD, Unions, Unions_tech, KSVD,Major}. However, in this case we have important prior knowledge that the basis $P$ is orthogonal 4-block diagonal. One way of using this knowledge is dividing the signals $X$ into 4 blocks corresponding to the 4 blocks of $P$, and estimating each block of $P$ from the relevant block of $X$ using the algorithm in Table~\ref{tab:UDL}, which is designed for learning an orthogonal basis.

\begin{table}[tbp]
\caption{dl algorithm for orthogonal dictionary}\label{tab:UDL}
\begin{center}
\begin{tabular}{|l|c|r|}
\hline
    \textbf{Inputs}\\
    $\bullet$ $X$ - training set\\
    $\bullet$ $k$ - sparsity level\\
\hline
    \textbf{Outputs}\\
    $\bullet$ $P$ - orthogonal dictionary\\
    $\bullet$ $S$ - sparse matrix\\
\hline
    \textbf{Algorithm}\\
    $\bullet$ Initiate $P=I$.\\
    $\bullet$ Repeat until a stoping criteria is reached:\\
        $\qquad \circ$ Fix $P$ and calculate $S=P^TX$.\\
        $\qquad \circ$ Keep only the $k$ highest (absolute value) elements \\
        $\qquad \quad$ in each column of $S$.\\
        $\qquad \circ$ Fix $S$, and calculate the SVD: $SX^T=U\Sigma V^T$.\\
        $\qquad \circ$ Update $P=VU^T$.\\
\hline
\end{tabular}
\end{center}
\end{table}

Due to this structure of $P$ and the sparsity of $S$ in each column of $X$ there are up to 12 nonzero elements. Therefore, the identity matrix $I$ was one of the bases in the finite set $\Psi$ that we used. Specifically, we used the same set $\Psi$ as in the simulations in Section~\ref{sec:finite}.
$X$ had about twice as many nonzero elements in each column compared to the real sparse matrix $S$, such that $X$ is $2k$-sparse under $I$. Therefore, we ran the F-BCS method with sparsity level of $2k$ instead of $k$.
Moreover, since $P$ is sparse itself we used $\Phi=I$ as the base dictionary in the sparse BCS method. It is easy to see that $k_p=2$.

Table~\ref{tab:compI} reports the average error of all five methods, calculated as in \eqref{eq:Err}.
As can be seen, the results of F-BCS are much worse than all the others. This can be expected since in this case $X$ is $2k$-sparse, so that the OMP reconstruction is not as good.
The error of the sparse BCS is also higher then the rest. The reason for this is that in order for the direct method of sparse BCS to work well the product $k_pk$ should be small relative to $n$. In this case this product is not small enough.
Note that though higher from the rest the errors of the sparse BCS and F-BCS are quite small.
We performed the same simulation with $k=3$ and then the error of sparse BCS was reduced to the level of the rest, but the error of F-BCS was still high.

The results of both the OBD-BCS algorithm and the CS with the estimated basis, which both did not use the knowledge of the basis $P$, are similar to those of the algorithm which used this knowledge. Thus, the prior knowledge of $P$ can be avoided. The advantage of OBD-BCS over the CS with the estimated basis is that it does not require any training set, and therefore can be used in applications where there is no access to any full signals but only to their measurements.

\begin{table}[tbp]
\caption{reconstruction error of different reconstruction algorithms}\label{tab:compI}
\begin{center}
\begin{tabular}{|l|c|r|}
\hline
Algorithm & Error \\
\hline
CS with the real $P$ & $10^{-5}\%$ \\
CS with $\hat{P}=P_{DL}$ & $10^{-5}\%$ \\
F-BCS & 0.522\% \\
Sparse BCS & 0.084\% \\
OBD-BCS & $10^{-5}\%$ \\
\hline
\end{tabular}
\end{center}
\end{table}


\section{Conclusions} \label{sec:conclusions}
We presented the problem of BCS which aims to solve CS problems without the prior knowledge of the sparsity basis of the signals.
Therefore, this work renders CS universal not only from the measurement process point of view, but also from the recovery point of view.

We presented three different constraints on the sparsity basis, that can be added to the BCS problem in order to guarantee the uniqueness of the solution to the BCS problem. Under each of these constraints we proved uniqueness conditions and proposed  simple methods to retrieve the solution.
All the proposed methods perform very well in simulations on synthetic data.
In fact, when $k$ is small enough and when enough signals are measured (only for the structural constraint case), the performance of our methods is similar to those of a standard CS which uses the real, though unknown in practice, sparsity basis.
We also demonstrated through simulations the advantage of BCS over CS with an estimated sparsity basis. The advantage of BCS is that it does not require any training set, and therefore can be used in applications where there is no access to any full signals but only to their measurements.

An interesting direction for future research is to examine more ways to assure uniqueness, beside the three presented here, and weaken the constraint on the basis.

\section{Acknowledgments}
The authors would like to thank Prof. David Malah and Mr. Moshe Mishali for fruitful discussions and helpful advice.

\renewcommand{\thetheorem}{\Alph{section}.\arabic{theorem}}
\renewcommand{\theequation}{\Alph{section}-\arabic{equation}}
\setcounter{section}{0}
\setcounter{equation}{0}
\setcounter{theorem}{0}

\addtocounter{section}{1}
\section*{Appendix \Alph{section}}
The following proves Lemma~\ref{lemma:PQ}. That is, if $P$ and $\hat{P}$ are both $2L$-block diagonal matrices, $A$ satisfies the conditions of Theorem~\ref{th:struct}, and $Q$ is a permutation matrix, then $A\hat{P}=APQ$ implies $\hat{P}=PQ$.

We begin this proof by proving that under the lemma's conditions $Q$ is necessarily block diagonal, after this is done the completion of the proof is straight forward.
For any $D=[D_1,...,D_L]\in \mathbb{R}^{n \times nL}$ such that $D_1,...,D_L\in \mathbb{R}^{n \times n}$ the permutation $DQ$ can yield three types of changes in $D$. It can mix the blocks of $D$, permute the order of the blocks of $D$, and permute the columns inside each block. $Q$ is $L$-block diagonal if and only if it permutes only the columns inside each block, but does not mix the blocks or change their outer order.

First we prove that $Q$ cannot mix the blocks of $D$. We denote by $Q_B$ the group of all block permutation matrices, which is the group of all the permutation matrices that keep all blocks together. That is, if $Q\in Q_B$ then when multiplying $DQ$ only the order of the blocks $D_1,...,D_L$ and the order of the columns inside the blocks change, but there is no mixture between the blocks.
After we prove that $Q\in Q_B$ we prove that $Q$ also cannot change the outer order of the blocks, and therefore must be block diagonal.
In order to prove that necessarily $Q\in Q_B$, we use the next two lemmas.

\begin{lemma}\label{lemma:D}
    If $D=[D_1,...,D_L]\in \mathbb{R}^{n\times nL}$ is a union of $L$ orthogonal bases, and $\sigma(D)=n+1$, then any set of $n$ orthogonal columns of $D$ are necessarily all from the same block of $D$.
\end{lemma}

\emph{Proof}:
Assume $\Gamma$ is a set of $n$ orthogonal columns from $D$. Denote $\Gamma=\Gamma_1\cup\Gamma_2$, where $\Gamma_1$ is the set of columns taken from $D_1$, and $\Gamma_2$ contains the rest of the columns in $\Gamma$. Without loss of generality assume the set $\Gamma_1$ is not empty.
Since both $D_1$ and $\Gamma$ are orthogonal bases of $\mathbb{R}^n$, the span of $\Gamma_2$ equals the span of the columns of $D_1$ which are not in $\Gamma$. Therefore, the set of columns $\Gamma_2\cup {d}$, where $d$ is any column from $D_1$ which is not in $\Gamma$, is either linearly dependent or empty.
However, the set $\Gamma_2\cup {d}$ contains at most $n$ columns, so that since $\sigma(D)=n+1$ this set cannot be linearly dependent. Therefore, $\Gamma_2$ is necessarily empty, such that all the columns of $\Gamma$ are from the same block of $D$.
\Qed

\begin{lemma}\label{lemma:Qb}
    Assume $D=[D_1,...,D_L]\in \mathbb{R}^{n\times nL}$ is a union of $L$ orthonormal bases, with $\sigma(D)=n+1$, and $\hat{D}=DQ$ for some permutation matrix $Q$. If $\hat{D}$ is also a union of $L$ orthonormal bases, then $Q\in Q_B$.
\end{lemma}

\emph{Proof}:
If there was a permutation $Q\notin Q_B$ such that $\hat{D}=DQ$, it would imply that $n$ columns of $D$, not all from the same block, form one of the orthogonal blocks of $\hat{D}$. However, according to Lemma~\ref{lemma:D} any $n$ orthogonal columns must be from the same block, and therefore $Q\in Q_B$ .
\Qed

We need to prove that the equality $A\hat{P}=APQ$ implies $\hat{P}=PQ$.
Denote the orthogonal blocks of $A$ by $A_i$ for $i=1,...,L$ and the orthogonal blocks of $P$ and $\hat{P}$ by $P^j$ and $\hat{P}^j$ respectively for $j=1,...,2L$. Also denote:
\begin{equation*}
\begin{split}
D=AP= \left[A_1 \left(\begin{array}{ccc} P^1 & \\& P^2\end{array}\right),...,A_L\left(\begin{array}{ccc} P^{2L-1} & \\& P^{2L}\end{array}\right)\right]\\
\hat{D}=A\hat{P}= \left[A_1 \left(\begin{array}{ccc} \hat{P}^1 & \\& \hat{P}^2 \end{array} \right),...,A_L \left(\begin{array}{ccc} \hat{P}^{2L-1} & \\& \hat{P}^{2L}\end{array}\right)\right]
\end{split}
\end{equation*}
which are both unions of $L$ orthogonal bases since $A_i$, $P^j$ and $\hat{P}^j$ are all orthogonal. Therefore, according to Lemma~\ref{lemma:Qb} $Q \in Q_B$.

Next we prove that $Q$ also cannot change the outer order of the blocks, and therefore must be $L$-block diagonal.
Assume by contradictions that $Q$ changes the outer order of the blocks of $D$. Without loss of generality we can assume this change is a switch between the first two blocks of $D$. That is,
\begin{equation*}
\begin{split}
    \hat{D}_1=D_2Q_2=A_2\left[\begin{array}{ccc} P^3 & \\ & P^4 \end{array} \right]Q_2 \\
    \hat{D}_2=D_1Q_1=A_1\left[\begin{array}{ccc} P^1 & \\ & P^2 \end{array} \right]Q_1
\end{split}
\end{equation*}
where $Q_1,Q_2$ are the corresponding sub-matrices of $Q$ which permute the columns inside the blocks $D_1,D_2$. In order to satisfy $\hat{D}=A\hat{P}$ we must have
\begin{equation*}
\begin{split}
\hat{D}_1=A_1\left[\begin{array}{ccc} \hat{P}^1 & \\ & \hat{P}^2 \end{array} \right]=A_2\left[\begin{array}{ccc} P^3 & \\ & P^4 \end{array} \right]Q_2 \\
\hat{D}_2=A_2\left[\begin{array}{ccc} \hat{P}^3 & \\ & \hat{P}^4 \end{array} \right]=A_1\left[\begin{array}{ccc} P^1 & \\ & P^2 \end{array} \right]Q_1.
\end{split}
\end{equation*}
Since $A_1$ and $A_2$ are orthogonal the above implies
\begin{equation}
\begin{split}\label{eq:blk_diag}
    \left[\begin{array}{ccc} \hat{P}^1 & \\ & \hat{P}^2 \end{array} \right]=A_1^TA_2\left[\begin{array}{ccc} P^3 & \\ & P^4 \end{array} \right]Q_2 \\
    \left[\begin{array}{ccc} \hat{P}^3 & \\ & \hat{P}^4 \end{array} \right]=A_2^TA_1\left[\begin{array}{ccc} P^1 & \\ & P^2 \end{array} \right]Q_1.
\end{split}
\end{equation}
If there is an orthogonal $2L$-block diagonal matrix $\hat{P}$ that satisfies \eqref{eq:blk_diag}, then in contradiction to Lemma~\ref{lemma:PQ} $\hat{P}\neq PQ$. However, \eqref{eq:blk_diag} implies:
\begin{equation*}
    A_1^TA_2=\left[\begin{array}{cc} \hat{P}^1 & \\ & \hat{P}^2 \end{array} \right]Q_2^T\left[\begin{array}{ccc} P^{3^T} & \\ & P^{4^T} \end{array} \right]=\left[\begin{array}{cc} R_1 & R_2\\ R_3 & R_4 \end{array} \right].
\end{equation*}
Due to the structure of the permutation matrix $Q_2$ and due to the orthogonality of the blocks of $P$ and $\hat{P}$, the ranks of $R_1,R_2,R_3,R_4$ must satisfy:
\begin{equation*}
\begin{split}
\lefteqn{\text{rank}(R_1)=\text{rank}(R_4)}\\
& \text{rank}(R_2)=\text{rank}(R_3)=\frac{n}{2}-\text{rank}(R_1).
\end{split}
\end{equation*}
Therefore, $A$ is necessarily inter block diagonal. However, according to the conditions of Theorem~\ref{th:struct} $A$ is not inter block diagonal, so that the contradictions assumption is incorrect and $Q$ cannot change the outer order of the blocks, such that $Q$ must be $L$-block diagonal.

Denote the diagonal blocks of $Q$ by $Q_i$ for $i=1,...,L$, such that:
\begin{equation*}
\begin{split}
\lefteqn{\hat{D}= \left[A_1 \left(\begin{array}{ccc} \hat{P}^1 & \\& \hat{P}^2 \end{array} \right),...,A_L \left(\begin{array}{ccc} \hat{P}^{2L-1} & \\& \hat{P}^{2L} \end{array} \right)\right]=}\\
& \left[A_1 \left(\begin{array}{ccc} P^1 & \\& P^2 \end{array} \right)Q_1,...,A_L \left(\begin{array}{ccc} P^{2L-1} & \\& P^{2L}\end{array}\right)Q_L\right].
\end{split}
\end{equation*}
Since all $A_i$ are orthogonal the above implies that for all $i=1,...,L$
\begin{equation*}
\begin{split}
\left[\begin{array}{ccc} \hat{P}^{2i-1}& \\& \hat{P}^{2i}\end{array}\right]=
\left[\begin{array}{ccc} P^{2i-1}& \\& P^{2i}\end{array}\right]Q_i,
\end{split}
\end{equation*}
such that $\hat{P}=PQ$.
\Qed

In fact the above proves not only that $Q$ is $L$-block diagonal, it is also $2L$-block diagonal.
Note that the extension of this proof to the case where $P$ and $\hat{P}$ have $ML$ blocks, for $M>2$, is trivial. However, if $P$ and $\hat{P}$ had $L$ blocks instead of $2L$, this proof would not work. That is since in this proof in order to eliminate solutions of the form of \eqref{eq:blk_diag} we use the 2-block diagonal structure of the matrices.
If there were only $L$ blocks, then beside the solution $\hat{P}=PQ$ there would have been another possibility, which is:
\begin{equation*}
\hat{P}=\left[\begin{array}{ccccc} A_1^TA_2P_2Q_2& & & &\\& A_2^TA_1P_1Q_1&&&\\&&P_3Q_3&&\\&&&\ddots&\\&&&&P_LQ_L \end{array}\right],
\end{equation*}
where $P_1,...P_L$ are the $L$ blocks of $P$ and $Q_1,...Q_L$ the the corresponding blocks of $Q$. Obviously in this case $\hat{P}\neq PQ$.

\addtocounter{section}{1}
\section*{Appendix \Alph{section}}
The following proves that if $A=[A_1,...,A_L]\in\mathbb{R}^{n\times nL}$ is a union of $L$ orthogonal bases, where each block is generated randomly from an i.i.d Gaussian distribution followed by a Gram-Schmidt process, then with probability~1 $\sigma(A)=n+1$ and $A$ is not inter-block diagonal (Definition~\ref{def:InterBlock}).
Multiplication by an orthogonal $P$ does not change the statistics, therefore if $\sigma(A)=n+1$ with probability~1, then also $\sigma(AP)=n+1$ with probability~1.
Therefore, such an $A$ satisfies the conditions of Theorem~\ref{th:struct} with probability~1.

We begin the proof by noting that we can look at the generation of each block of $A$ as follows. The first column $a_1$ is generated randomly from $\mathbb{R}^{n}$. The second column $a_2$ is generated randomly from the $n-1$ dimensional space orthogonal to $a_1$. the column $a_3$ is generated randomly from the $n-2$ dimensional space orthogonal to the span of $\{a_1,a_2\}$, and similarly any $a_i$ is generated randomly from the space orthogonal to the span of all previous columns, whose dimension is $n-i+1$.
We start by proving $\sigma(A)=n+1$. This proof uses the next lemma.

\begin{lemma}\label{lemma:gamma}
Assume $G\in\mathbb{R}^{n\times n}$ is generated as an i.i.d Gaussian matrix followed by a Gram-Schmidt process, and $U$ is a given space of dimension $d$. If $d<n$ then with probability~1 non of the columns of $G$ are in $U$.
\end{lemma}

\emph{Proof}:
Denote the columns of $G$ by $g_i$ for $i=1,...,n$. Since $d<n$ the space $U$ has zero volume in $\mathbb{R}^{n}$. $g_1$ is generated randomly from $\mathbb{R}^{n}$, and therefore with probability~1 $g_1$ is not in $U$.
For any other $1<i\leq n$, $g_i$ is generated randomly from $G_i$, which is the space orthogonal to the $i-1$ previous columns in $G$. $G_i$ dimension is $d_i=n-i+1$.
In this case we need to look at the probability to generate $g_i$ in the intersection $U\cap G_i$.
If $d<d_i$ then obviously this intersection has zero volume in $G_i$, so that $g_i$ is not in $U$ with probability~1.
Furthermore, if $d\geq d_i$ then due to the randomness of the columns of $G$, $G_i$ is not entirely contained in $U$ with probability~1. Therefore, here too $U\cap G_i$ has zero volume in $G_i$, such that $g_i$ is not in $U$ with probability~1.
\Qed \\[0.5cm]
\indent Assume $\Gamma$ is a set of $\sigma(A)$ linearly dependent columns from $A$.
Denote $\Gamma=\Gamma_1\cup\Gamma_2$, where $\Gamma_1$ is the subset of $\Gamma$ which contains only the columns taken from the block $A_1$, and $\Gamma_2$ are the rest of the columns in $\Gamma$. Without loss of generality assume $\Gamma_1$ is not empty. Moreover, since $A_1$ is orthogonal $\Gamma_1$ is also orthogonal, such that in order for $\Gamma$ to be linearly dependent $\Gamma_2$ also cannot be empty.\\
\indent Any $n+1$ columns from $A$ are linearly dependent such that $\sigma(A)\leq n+1$. Therefore, $|\Gamma|\leq n+1$ so that $|\Gamma_1|,|\Gamma_2|\leq n$.
If $|\Gamma_1|=n$ or $|\Gamma_2|=n$ then necessarily $\sigma(A)=|\Gamma|=n+1$.
Assume by contradiction that $\sigma(A)=|\Gamma|\leq n$, such that $|\Gamma_1|<n$ and $|\Gamma_2|\leq n-|\Gamma_1|$.
If $|\Gamma_1|$ contains only one column, denoted by $\gamma_1$, then since $\Gamma$ is linearly dependent $\gamma_1$ must be in the span of $\Gamma_2$. However, the dimension of this span is at most $|\Gamma_2|\leq n-1$, such that according to Lemma~\ref{lemma:gamma} the probability for this is zero.
If $\Gamma_1$ contains only two columns, denoted by $\gamma_1,\gamma_2$, then $\gamma_2$ must be in the span of $\Gamma_2\cup \gamma_1$. However, the dimension of this space is at most $|\Gamma_2|+1\leq n-1$, such that according to Lemma~\ref{lemma:gamma} the probability for this is again zero.
We can keep increasing the cardinality of $\Gamma_1$ and as long as $|\Gamma|\leq n$ the probability for $\Gamma$ to be linearly dependent will be zero. Therefore, the contradiction assumption is incorrect with probability~1, so that $\sigma(A)=|\Gamma|=n+1$ with probability~1.

Next we need to prove that $A$ is not inter-block diagonal. Denote for any pair of indices $i\neq j$:
\begin{equation}
A_i^TA_j=\left[\begin{array}{cc}R_1&R_2\\R_3&R_4\end{array}\right].
\end{equation}
For $A$ to be inter block diagonal there should be a pair $i\neq j$ for which:
\begin{equation}
\begin{split}
\lefteqn{\text{rank}(R_1)=\text{rank}(R_4)}\\
& \text{rank}(R_2)=\text{rank}(R_3)=\frac{n}{2}-\text{rank}(R_1).
\end{split}
\end{equation}
However, due to the randomness of $A_i,A_j$ the blocks $R_1,R_2,R_3,R_4$ all have full rank with probability~1.
So that $\text{rank}(R_1)=\text{rank}(R_2)=\frac{n}{2}$ and $\text{rank}(R_2)\neq \frac{n}{2}-\text{rank}(R_1)$. Therefore, $A$ is not inter block diagonal with probability~1.

\addtocounter{section}{1}
\section*{Appendix \Alph{section}}
Assume $A\in\mathbb{R}^{\frac{m}{L}\times m}$ is a union of $L$ random orthogonal bases and $P\in\mathbb{R}^{m\times m}$ is an orthogonal $2L$-block diagonal matrix. Denote $\tilde{D}=APQ$ where $Q$ is some unknown signed permutation matrix. We prove here that there are $[(\frac{m}{2L})!]^{2L}$ different permutation matrices $Q_D$ such that $\tilde{D}Q_D=A\hat{P}$, where $\hat{P}$ is an orthogonal $2L$-block diagonal matrix. Without loss of generality we can assume $Q=I$, therefore we need to refer to $APQ_D=A\hat{P}$. According to Lemma~\ref{lemma:PQ} this implies $PQ_D=\hat{P}$. Since both $P$ and $\hat{P}$ are $2L$-block diagonal $Q_D$ must be too, and the size of its blocks is $\frac{m}{2L}\times\frac{m}{2L}$. $Q_D$ is a permutation matrix, therefore each of its blocks is a permutation of the identity matrix of size $\frac{m}{2L}$. Thus, there are only $(\frac{m}{2L})!$ different possibilities for each block. There are $2L$ blocks such that the total number of possible $Q_D$'s is $[(\frac{m}{2L})!]^{2L}$.

\bibliographystyle{IEEEbib}
\bibliography{bib}
\end{document}